\documentstyle[preprint,eqsecnum,aps]{revtex}
\begin{document}
\draft
\preprint{IPNO/TH 94-5}
\title{The two-fermion relativistic wave equations\\of Constraint Theory in
the Pauli-Schr\"odinger form}
\author{J. Mourad}
\address{Laboratoire de Physique Th\'eorique et Hautes Energies\thanks
{Laboratoire associ\'e au CNRS.},\\Universit\'e Paris XI, B\^at. 211,\\
F-91405 Orsay Cedex, France}
\author{H. Sazdjian}
\address{Division de Physique Th\'eorique\thanks{Unit\'e de Recherche
des Universit\'es Paris 11 et Paris 6 associ\'ee au CNRS.}, Institut de
Physique Nucl\'eaire,\\ Universit\'e Paris XI,\\F-91406 Orsay Cedex, France\\
\vspace {7 cm}
}
\maketitle
{\flushleft LPTHE 94/16}
\pagebreak

\begin{abstract}

The two-fermion relativistic wave equations of Constraint Theory are reduced,
after expressing the components of the $4\times 4$ matrix wave
function in terms of one of the $2\times 2$ components, to a single equation
of the Pauli-Schr\"odinger type, valid for all sectors of quantum numbers.
The potentials that are present belong to the general classes of scalar,
pseudoscalar and vector interactions and are calculable in perturbation
theory from Feynman diagrams.
In the limit when one of the masses becomes infinite, the equation reduces
to the two-component form of the one-particle Dirac equation with
external static potentials.
The Hamiltonian, to order $1/c^2$, reproduces
most of the known theoretical results obtained by other methods. The gauge
invariance of the wave equation is checked, to that order,
in the case of QED.
The role of the c.m. energy dependence of the relativistic interquark
confining potential is emphasized and the structure of the Hamiltonian,
to order $1/c^2$, corresponding to confining scalar potentials,
is displayed.

\end{abstract}

\pacs{PACS numbers : 11.10.Qr, 11.10.St, 12.40.Qq.}
\narrowtext

\section {Introduction}

The use of the manifestly covariant formalism with constraints \cite{ll} in the
two-body problem \cite{cva1,s1} leads to a Poincar\'e invariant description
of the dynamics of the system with the correct number of degrees of freedom.
Furthermore, the potentials that appear in the corresponding wave equations
are calculable, in perturbation theory, in terms of the kernel of the
Bethe-Salpeter equation, and therefore allow one to deal with quantum field
theoretic problems.\par
For two spin-0 particles, the system is described by two independent wave
equations, which are generalizations of the individual Klein-Gordon equation
of each particle, including now the mutual interaction potential. The
compatibility condition of the two equations imposes certain restrictions on
the structure of the potential and leads, in a covariant form, to an
elimination of the relative energy variable. One then ends up with a final,
manifestly covariant, three-dimensional eigenvalue equation that describes
the relative motion of the two particles. This equation is very similar in
form, except for kinematic energy dependent factors, to the Schr\"odinger
(or Klein-Gordon) equation : it is a second order differential equation in the
three spacelike coordinates (for local potentials) and therefore the usual
techniques of nonrelativistic quantum mechanics are applicable to it.\par
For two spin-1/2 particles, the system is described by two independent
Dirac type equations. Again, the compatibility condition imposes
restrictions on the stucture of the potentials and eliminates the relative
energy variable; but, because of the presence of the Dirac matrices, the
reduction to a final eigenvalue equation is not straightforward. Except for
certain classes of interaction, as the pseudoscalar interaction and the vector
interaction without temporal components, the reduction process is rather
complicated and dependent on the way of eliminating the components of the
spinor wave function in terms of one of them. Up to now, no single
Pauli-Schr\"odinger type equation was obtained from this procedure, capable
of describing general cases of interaction and all sectors of quantum numbers,
although it was possible to obtain two coupled Pauli-Schr\"odinger type
equations for the sectors of quantum numbers with $\ell =\vert j\pm 1\vert$ and
single equations for the sectors with $\ell =j$ \cite{cbwva}.\par
The purpose of the present paper is to show that it is possible, for general
combinations of scalar, pseudoscalar and vector interactions, the latter being
considered in arbitrary covariant gauges, to reduce the wave equations
describing two spin-1/2 particle systems to a single Pauli-Schr\"odinger type
equation, valid for all sectors of quantum numbers. This is achieved by
decomposing the $4\times 4$ spinor wave function (of a fermion-antifermion
system, say) along $2\times 2$ components defined, in the c.m. frame, by the
basis of the matrices $1,\ \gamma _0,\ \gamma _5$ and $\gamma _0 \gamma _5$,
rather than the usual basis $(1\pm \gamma _0)$ and $(1\pm \gamma _0)\gamma _5$.
\par
The above reduction is obtained in the case of quasilocal approximations of the
interaction potentials, i.e., for potentials that are functions of the
relative coordinates and, eventually, of the c.m. energy. This does not cover
the more general case of integral operators (in the three spacelike
coordinates), but it is generally admitted that a local expression of the
potential may provide a valid zeroth order approximation of the total
interaction, from which one may develop perturbative calculations for the
nonlocal effects. Furthermore, an appropriate c.m. energy dependence of
the potential takes into account the leading contribution of the nonlocal
effects.\par
The final Pauli-Schr\"odinger type equation that is obtained provides
explicit and  simple means of controlling the relativistic effects of each
type of interaction.\par
The contents of the paper are summarized as follows. In Sec. II, we review
the main aspects of the relativistic wave equations of Constraint Theory for
a fermion-antifermion system. The calculation of the potentials from
one-particle exchange Feynman diagrams is presented in Appendix A.\par
In Sec. III, we reduce these wave equations to a single two-body
Pauli-Schr\"odinger type equation, for the case of a general combination of
scalar, pseudoscalar and vector interactions. Some technical details of the
reduction process and useful formulas for the $\gamma$-matrices and the spin
and orbital angular momentum operators are presented in Appendix B.\par
In Sec. IV, we study the limit of the above equation when one of the particles
becomes infinitely massive. It reduces to the Pauli-Schr\"odinger form of the
one-particle Dirac equation in the presence of the static potential created
by the heavy particle.
In Appendix C, we show that, for the class of ladder and crossed ladder type
diagrams, it is only the one-particle exchange diagram which contributes in
this limit to the expression of the potential.\par
Section V deals with the nonrelativistic limit, to order $1/c^2$, of the above
two-body Pauli-Schr\"odinger type equation. Most of the known theoretical
results obtained by other methods are reproduced. In the case of QED, its gauge
invariance, to order $1/c^2$, is checked by considering the photon
propagator in arbitrary covariant gauges.
The role of the c.m. energy dependence of the relativistic potentials
is emphasized and the structure of the Hamiltonian, for confining
scalar potentials, ready for comparisons with quantum field theoretic
results, is displayed.\par
Summary and concluding remarks follow in Sec. VI.\par
\pagebreak

\section {The two-body relativistic wave equations \protect \\
of Constraint Theory}

For a system of two spin-1/2 particles, composed of one fermion of mass $m_1$
and one antifermion of mass $m_2$, say, Constraint Theory imposes two
independent wave equations, which are generalizations of the Dirac equation
satisfied by each particle in the free case. These wave equations have not a
unique form and can be modified by wave function transformations, but a
convenient form, where several properties can be easily read off, is the
following one \cite{s1} :
\begin{mathletters}
\begin{eqnarray}
\big (\ \gamma _1.p_1 - m_1\big )\ \widetilde \Psi \ =\
\big (-\gamma _2.p_2 + m_2\big )\ \widetilde V\ \widetilde \Psi \ ,\\
\big (-\gamma _2.p_2 - m_2\big )\ \widetilde \Psi \ =\
\big (\ \gamma _1.p_1 + m_1\big )\ \widetilde V\ \widetilde \Psi \ .
\end{eqnarray}
\end{mathletters}
Here, $\widetilde \Psi$ is a sixteen-component spinor wave function of rank
two and is represented as a $4\times 4$ matrix :
\begin{equation}
\widetilde \Psi \ =\ \widetilde \Psi _{\alpha _1,\alpha _2} (x_1,x_2)\ \ \ \
(\alpha _1,\alpha _2=1,\ldots ,4)\ ,
\end{equation}
where $\alpha _1 (\alpha _2)$ refers to the spinor index of particle 1(2).
$\gamma _1$ is the Dirac matrix $\gamma $ acting in the subspace of the
spinor of particle 1 (index $\alpha _1$); it acts on $\widetilde \Psi $
from the left. $\gamma _2$ is the Dirac matrix acting in the subspace of
the spinor of particle 2 (index $\alpha _2$); it acts on $\widetilde \Psi $
from the right; this is also the case of products of $\gamma _2$ matrices,
which act on $\widetilde \Psi$ from the right in the reverse order :
\begin{eqnarray}
\gamma _{1\mu} \widetilde \Psi & \equiv &(\gamma _\mu )_{\alpha _1 \beta _1}
\widetilde \Psi _{\beta _1 \alpha _2}\ ,\ \ \
\gamma _{2\mu} \widetilde \Psi \equiv \widetilde \Psi _{\alpha _1 \beta _2}
(\gamma _\mu )_{\beta _2 \alpha _2}\ ,\nonumber \\
\gamma _{2\mu}\gamma _{2\nu} \widetilde \Psi & \equiv &
\widetilde \Psi _{\alpha _1 \beta _2} (\gamma _\nu \gamma _\mu)_
{\beta _2 \alpha _2}\ ,\ \ \
\sigma _{a\alpha \beta} = \frac {1}{2i} \big [ \gamma _{a\alpha },
\gamma _{a\beta } \big ]\ \ \ (a=1,2)\ .
\end{eqnarray}
In Eqs. (2.1) $p_1$ and $p_2$ represent the momentum operators of particles
1 and 2, respectively. $\widetilde V$ is a Poincar\'e invariant potential.\par
Equations (2.1) must be compatible among themselves. This is enforced in
two steps. First, one multiplies Eq. (2.1a) by $(\gamma _1.p_1 + m_1)$
and Eq. (2.1b) by $(-\gamma _2.p_2 + m_2)$ and subtracts the two equations
from each other. This yields the constraint :
\begin{equation}
\big [\ (p_1^2 - p_2^2) - (m_1^2 - m_2^2)\ \big ]\ \widetilde \Psi \ =\ 0\ ,
\end{equation}
which allows one to eliminate the relative energy variable in a covariant
form. For eigenfunctions of the total momentum operator $P$, the solution
of Eq. (2.4) is :
\begin{equation}
\widetilde \Psi \ =\ e^{\displaystyle -iP.X}\
e^{\displaystyle -i(m_1^2 - m_2^2) P.x/(2P^2)}\
\widetilde \psi (x^T)\ ,
\end{equation}
where we have used notations from the following definitions :
\begin{eqnarray}
P &=& p_1 + p_2\ ,\ \ \ p = \frac {1}{2} (p_1 - p_2)\ ,\ \ \
M = m_1 + m_2\ , \nonumber \\
X &=& \frac {1}{2} (x_1 + x_2)\ ,\ \ \ x = x_1 - x_2\ ,\ \ \
\mu = \frac {m_1 m_2}{M}\ .
\end{eqnarray}
We also define transverse and longitudinal components of four-vectors with
respect to the total momentum $P$ :
\begin{eqnarray}
q_\mu ^T &=& q_\mu - \frac {(q.P)}{P^2} P_\mu\ ,\ \ \
q_\mu ^L = (q.\hat P) \hat P_\mu \ ,\ \ \ \hat P_\mu = P_\mu /\sqrt {P^2}\ ,
\nonumber \\
q_L &=& q.\hat P\ ,\ \ \ P_L = \sqrt {P^2}\ .
\end{eqnarray}
This decomposition is manifestly covariant. In the c.m. frame the transverse
components reduce to the three spacelike components, while the longitudinal
component reduces to the timelike component of the corresponding four-vector.
(Note that $x^{T2} = - {\bf x}^2$ in the c.m. frame.)\par
Second, one multiplies Eq. (2.1a) by $(-\gamma _2.p_2 - m_2)$ and Eq. (2.1b)
by $(\gamma _1.p_1 - m_1)$ and subtracts the two equations from each other.
Taking into account the constraint (2.4), one ends up with the new
constraint :
\begin{equation}
\big [\ p_1^2 - p_2^2\ , \widetilde V\ \big ]\ \widetilde \Psi \ =\ 0\ ,
\end{equation}
which concerns the potential $\widetilde V$. By noticing that
$p_1^2 - p_2^2 = 2P_L p_L$, this equation is satisfied by demanding that
$\widetilde V$, which is Poincar\'e invariant, be independent of the
longitudinal relative coordinate $x_L$ :
\begin{equation}
\widetilde V\ =\ \widetilde V(x^T,\ P_L,\ p^T,\ \gamma _1,\ \gamma _2)\ .
\end{equation}
\par
No other constraints are found. Equations (2.5) and (2.9) show that the
internal dynamics of the system is three-dimensional, besides the spin degrees
of freedom, described by the three-dimensional transverse coordinate $x^T$.
\par
To find the connection of Eqs. (2.1) with the Bethe-Salpeter equation
\cite{sbgml,n}, one projects the latter on the constraint hypersurface (2.4)
by expanding the right-hand side of the equation on that hypersurface
\cite{s2}. One ends up precisely with Eqs. (2.1), where now the potential
$\widetilde V$ is calculable, in perturbation theory, in terms of Feynman
diagrams.\par
The relationship between the potential $\widetilde V$ and Feynman diagrams
can be summarized by the following
Lippmann-Schwinger-Quasipotential type
\cite{ltlttk,bs,g,pl,f,fh,t,lcl} equation (momentum integrations, as well
as the total momentum $P$ are not explicitly written) :
\begin{eqnarray}
&\ &\widetilde V\ -\ \widetilde T\ -\ \widetilde V G_0 \widetilde T\ =\ 0\ ,\\
&\ &\widetilde T(p^T,p'^T)\ \equiv \ \frac {i}{2P_L} \bigg [\ T(p,p')\
\bigg ]_{C(p),C(p')}\ ,\nonumber
\end{eqnarray}
where :\\
i) $T$ is the off-mass shell fermion-antifermion scattering
amplitude (amputated four-point connected Green's function);\\
ii) $C$ is the constraint (2.4) :
\begin{equation}
C(p)\ \equiv \ (p_1^2 - p_2^2) - (m_1^2 - m_2^2)\ =\ 2P_Lp_L - (m_1^2 - m_2^2)
\ \approx \ 0\ ;
\end{equation}
in Eq. (2.10) the external momenta of the amplitude $T$ are submitted to
the constraint $C$;\\
iii) $G_0$ is defined as :
\begin{equation}
G_0(p_1,p_2)\ =\ \widetilde S_1(p_1)\ \widetilde S_2(-p_2)\ H_0\ ,
\end{equation}
where $\widetilde S_1$ and $\widetilde S_2$ are the propagators of the
two fermions, respectively, in the presence of the constraint
(2.11), and $H_0$ is the Klein-Gordon operator, also in the presence of
the constraint (2.11) :
\widetext
\begin{equation}
H_0\ =\ (p_1^2 - m_1^2) \bigg \vert _C \ =\ (p_2^2 - m_2^2) \bigg \vert _C \ =\
\frac{P^2}{4} - \frac{1}{2}(m_1^2 + m_2^2) + \frac {(m_1^2 - m_2^2)^2}
{4P^2} + p^{T2}\ .
\end{equation}
\narrowtext
\par
The series expansion of Eq. (2.10) involves not only the usual Feynman
diagrams of the amplitude $T$, but also additional diagrams, which we call
``constraint diagrams'', coming from the iteration of the $\widetilde V G_0$
term; they are obtained from the reducible diagrams by the replacement in
a box diagram of the fermion and antifermion propagators by the factor
$G_0$ (2.12) together with the $\delta$-function of the constraint (2.11).
It is the presence of these diagrams, as well as of the c.m. energy factor in
the function $\delta (C)$, that cures the Bethe-Salpeter equation of some
of its deficiencies \cite{s2}. (See also Appendix C.)
[We have included the c.m. energy factor $1/(2P_L)$ in the definition of
$\widetilde T$, and therefore, by imposing on $G_0$ the constraint (2.11),
the integrations relative to $G_0$ in Eq. (2.10) may be considered as
three-dimensional.]\par
In general $\widetilde V$ is an integral operator in $x^T$. However, in
one-particle exchange diagrams $\widetilde V$ turns out to be a function of
$x^T$ and $P_L$, with definite dependences on the $\gamma$-matrices. It
would also be meaningful to approximate, as a zeroth order approximation,
multiparticle exchange contributions by appropriate local functions and
thus to use for $\widetilde V$ quasilocal expressions (in $x^T$ and $P_L$).
Furthermore, in perturbation theory, the c.m. energy ($P_L$) dependence of
the potential takes into account the leading contribution of the nonlocal
effects. Therefore, we shall confine ourselves in the present paper to
quasilocal expressions of $\widetilde V$. The explicit expressions of
$\widetilde V$ in lowest order of perturbation theory, for the scalar,
pseudoscalar and vector interactions, are presented in Appendix A.\par
The scalar product associated with the solutions of Eqs. (2.1) can be
constructed by means of tensor currents of rank two, satisfying two
independent conservation laws, with respect to $x_1$ and $x_2$ \cite{s1}.
Another way of proceeding consists in using the integral equation of the
Green's function \cite{f,lcl}. From either method, one finds for the norm of
$\widetilde \psi$ [Eq. (2.5)] the formula (in the c.m. frame) :
\widetext
\begin{equation}
\int d^3{\bf x}\ Tr \bigg \{ \widetilde \psi ^{\dagger }\ \big [1 - \widetilde
V^{\dagger } \widetilde V + 4\gamma _{10} \gamma _{20} P_0^2 \frac
{\partial \widetilde V}{\partial P^2} \big ]\ \widetilde \psi \bigg \}\ =\
2P_0\ .
\end{equation}
\narrowtext
(The potentials $\widetilde V$ that are considered in the sequel commute
with the matrix product $\gamma _{1L} \gamma _{2L}$.)\par
This formula shows that for energy independent potentials (in the c.m.
frame) the norm of $\widetilde \psi$ is not positive definite. In order
to reach such a situation, it is sufficient that the potential $\widetilde V$
satisfy the inequality
\begin{equation}
\frac{1}{4} Tr (\widetilde V^{\dagger } \widetilde V) \ < \ 1\ ,
\end{equation}
which, in turn imposes restrictions on the high order (multiparticle
exchange) contributions to $\widetilde V$. Once condition (2.15) is
satisfied, one is allowed to make the wave function transformation
\begin{equation}
\widetilde \Psi\ =\ \big [ 1 - \widetilde V^{\dagger } \widetilde V \big ]
^{-\frac {1}{2}}\ \Psi
\end{equation}
and to reach a representation of the wave equations (2.1), where the norm
for energy independent potentials is the free norm.\par
In this connection, it was observed by Crater and Van Alstine \cite{cva2}
that the parametrization
\begin{equation}
\widetilde V \ =\ \tanh V
\end{equation}
satisfies condition (2.15) and also has the property that in the new
representation (2.17) one obtains familiar Dirac type wave equations, where
each particle appears as placed in the external potential created by the
other particle, the latter potential having the same tensor nature as
potential $V$ of Eq. (2.17).\par
We shall henceforth adopt parametrization (2.17) for potential $\widetilde V$
and, according to Eq. (2.16), shall introduce the wave function
transformation :
\begin{equation}
\widetilde \Psi \ =\ \big (\cosh V\big )\ \Psi\ ,
\end{equation}
$V$ being assumed to be hermitian. The norm of the new wave function
$\Psi$ then becomes (in the c.m. frame) :
\widetext
\begin{equation}
\int d^3{\bf x}\ Tr \bigg \{ \psi^{\dagger }\ \big [ 1 + 4\gamma _{10}
\gamma _{20} P_0^2 \frac {\partial V}{\partial P^2} \big ]\ \psi \bigg \}\ =\
2P_0\ .
\end{equation}
\narrowtext
(The relationship between $\Psi$ and $\psi$ is the same as in Eq. (2.5).)\par
The new potential $V$ will be chosen in the major part of this paper as a
general combination of scalar, pseudoscalar and vector potentials :
\widetext
\begin{equation}
V\ =\ V_1 + \gamma _{15} \gamma _{25} V_3 + \gamma _1^{\mu} \gamma _2^{\nu}
\ \big (\ g_{\mu \nu }^{LL} V_2 + g_{\mu \nu }^{TT} U_4 + \frac {x_{\mu }^T
x_{\nu }^T}{x^{T2}} T_4\ \big )\ ,
\end{equation}
\narrowtext
where the potentials $V_1,\ V_2,$ etc., are functions of $x^{T2}$ and $P_L$.
(The index notations of the potentials will be clarified in Sec. III.)
The dependence of the vector potentials on the gauge chosen for the photon
propagator in lowest order of perturbation theory is displayed in Appendix A.
Also notice that the difference between potentials $\widetilde V$ and $V$
shows up starting from third order diagrams.\par
When parametrization (2.17) is adopted and the change of function (2.18) is
used, Eqs. (2.1) take the form :
\begin{mathletters}
\begin{eqnarray}
(\ \gamma _1.p_1 - m_1)\ \cosh V \ \psi \ =\ (-\gamma _2.p_2 + m_2)
\ \sinh V \ \psi \ ,\\
(-\gamma _2.p_2 - m_2)\ \cosh V \ \psi \ =\ (\ \gamma _1.p_1 + m_1)
\ \sinh V \ \psi \ ,
\end{eqnarray}
\end{mathletters}
where, according to Eq. (2.5), $\psi (x^T)$ is the internal part of the
wave function $\Psi $. Also, because of the constraint (2.4) (or (2.11)),
the longitudinal components of the momentum operators are replaced by their
eigenvalues determined in terms of $P_L$ [Eq. (2.7)] and the masses :
\begin{mathletters}
\begin{eqnarray}
p_{1L}\ =\ \frac {P_L}{2} + \frac {(m_1^2 - m_2^2)}{2P_L}\ ,\\
p_{2L}\ =\ \frac {P_L}{2} - \frac {(m_1^2 - m_2^2)}{2P_L}\ .
\end{eqnarray}
\end{mathletters}
\par
By using the expression of $V$ [Eq. (2.20)] and the properties of the
$\gamma $-matrices (see Appendix B), one can bring the functions $\cosh V$
and $\sinh V$ to the left of the Dirac operators in Eqs. (2.21). If one
defines $\overline V \equiv 2V_1 - V$, then, upon multiplying Eq. (2.21a) by
$\cosh \overline V$ and Eq. (2.21b) by $-\sinh \overline V$ and subtracting
the two equations from each other, one obtains a Dirac type equation for
particle 1 as if it was placed in the presence of external potentials
created by particle 2 :
\widetext
\begin{eqnarray}
\bigg \{  &\big [ & \frac {P_L}{2} e^{\displaystyle 2V_2} +
\frac {(m_1^2 - m_2^2)}{2P_L}
e^{\displaystyle -2V_2}\ \big ]\ \gamma _{1L}  -
\frac {M}{2} e^{\displaystyle 2V_1}
- \frac {(m_1^2 - m_2^2)}{2M} e^{\displaystyle -2V_1} \nonumber \\
&+& e^{\displaystyle -2U_4}\ \bigg [ \ \gamma _1^T.p^T + \frac
{i\hbar}{2x^{T2}}
(e^{\displaystyle -2T_4} - 1)\ (2\gamma _1^T.x^T +
i\gamma _1^{T\alpha } \sigma _{2\alpha \beta } ^{TT} x^{T\beta })
+ (e^{\displaystyle -2T_4} - 1) \frac {\gamma _1^T.x^T}{x^{T2}}
x^T.p^T\nonumber \\
&-& 2i\hbar e^{\displaystyle -2T_4} \gamma _2^T.x^T\
\big (\dot V_1 + \gamma _{1L}
\gamma _{2L} \dot V_2 + \gamma _{15} \gamma _{25} \dot V_3 + \gamma _1^T.
\gamma _2^T \dot U_4 + \frac {\gamma _1^T.x^T \gamma _2^T.x^T}{x^{T2}} \dot T_4
\big )\ \bigg ]\ \bigg \}\ \psi \ =\ 0\ ,\nonumber \\
\
\end{eqnarray}
\narrowtext
where we have defined :
\begin{equation}
\dot F \ \equiv \ \frac {\partial F}{\partial x^{T2}}\ .
\end{equation}
\par
The Dirac type equation for particle 2 can be obtained from Eq. (2.23) by the
replacements : $p_1 \leftrightarrow -p_2,\ x \rightarrow x,\ m_1
\leftrightarrow m_2,\ \gamma _1 \leftrightarrow \gamma _2$ .\par
Equation (2.23) shows that the scalar potential acts essentially on the mass
terms, the timelike vector potential acts on the energy terms, the spacelike
vector potentials act on the transverse momentum dependent terms, while
the pseudoscalar potential acts only through spin dependent terms. These
results are consistent with the interpretation of Eq. (2.23) from the
external potential standpoint. In particular, we did not find, in passing
from Eqs. (2.21) to Eq. (2.23) new types of interaction. This is not a
trivial property. Had we used a different parametrization than that provided
by Eq. (2.17), for instance by decomposing $\widetilde V$, rather than $V$,
according to Eq. (2.20), we would have found that the original vector
potentials induced in the analogue of Eq. (2.23) additional axial vector
terms.\par
\pagebreak

\section{Reduction of the wave equations \protect \\
to the Pauli-Schr\"odinger form}

In order to bring Eqs. (2.21) into a more transparent form, we decompose the
$4\times 4$ matrix wave function on the basis of the matrices
$1,\ \gamma _L,\ \gamma _5$ and $\gamma _L \gamma _5$ by defining $2\times 2$
matrix components :
\widetext
\begin{equation}
\psi \ =\ \psi _1 + \gamma _L \psi _2 + \gamma _5 \psi _3 + \gamma _L
\gamma _5 \psi _4 \ \equiv \ \sum _{i=1}^{4} \ \Gamma _i \psi _i \ .
\end{equation}
\narrowtext
\par
Equations (2.21) are then projected in the above subspaces. We shall first
consider the case of more general potentials than those introduced in
Eqs. (2.20) and develop our method of calculation for this general case.
They have the property of being functions of products of $\gamma _1$ and
$\gamma _2$ matrices in equal number. This feature is obvious for one
particle exchange diagrams (for parity conserving interactions), where
the matrices $\gamma _1$ and $\gamma _2$ couple symmetrically to the
propagator of the exchanged particle. However, when general vertex
corrections are considered, one finds expressions in which some terms may
not have the same number of $\gamma _1$ and $\gamma _2$ matrices,
although globally the total expressions are symmetric in the exchanges
$1\leftrightarrow 2$. Such terms will not be considered here. Then, the
most general (parity and time reversal invariant) potential we may
consider has the following decomposition on the basis (3.1) :
\begin{equation}
V\ =\ v_1 + \gamma _{1L} \gamma _{2L}\ v_2 + \gamma _{15} \gamma _{25}\ v_3
+ \gamma _{1L} \gamma _{15} \gamma _{2L} \gamma _{25}\ v_4 \ ,
\end{equation}
where the potentials $v_i \ (i=1,\ldots ,4)$ may still have spin
dependences.\par
Spin operators, which act in the $2\times 2$ component subspaces, are
defined by means of the Pauli-Lubanski operators :
\widetext
\begin{eqnarray}
W_{1S\alpha }\ &=&\ -\frac {\hbar }{4} \epsilon _{\alpha \beta \mu \nu }
P^{\beta } \sigma _1^{\mu \nu }\ ,\ \
W_{2S\alpha }\ =\ -\frac {\hbar}{4} \epsilon _{\alpha \beta \mu \nu }
P^{\beta } \sigma _2^{\mu \nu }\ \ \ (\epsilon _{0123} = +1)\ ,\nonumber \\
W_{1S}^2 \ &=&\ W_{2S}^2 \ =\ -\frac {3}{4} \hbar ^2 P^2\ ,\ \ \
W_S\ =\ W_{1S} + W_{2S}\ .
\end{eqnarray}
\narrowtext
They also satisfy the relations :
\begin{equation}
\gamma _{1L} W_{1S\alpha }\ =\ \frac {\hbar P_L}{2} \gamma _{1\alpha }^T
\gamma _{15}\ ,\ \ \ \gamma _{2L} W_{2S\alpha }\ =\ \frac {\hbar P_L}{2}
\gamma _{2\alpha }^T \gamma _{25}\ .
\end{equation}
(Note that the $W$'s are transverse vectors.)\par
We introduce the operators :
\begin{mathletters}
\begin{eqnarray}
w\ &=&\ (\frac {2}{\hbar P_L})^2\ W_{1S}.W_{2S}\
\stackrel {c.m.}{\longrightarrow }\
-\frac{4}{\hbar ^2} {\bf s_1.s_2}\ ,\\
w_{12}\ &=&\ (\frac {2}{\hbar P_L})^2\ \frac {W_{1S}.x^T W_{2S}.x^T}{x^{T2}}\
\stackrel {c.m.}{\longrightarrow }\
-\frac{4}{\hbar ^2} {\bf \frac {(s_1.x)(s_2.x)}{x^2}}\ \equiv \ -S_{12}\ .
\end{eqnarray}
\end{mathletters}
In the preceding relations we also displayed the corresponding c.m.
expressions in terms of the usual spin operators.\par
Then, the potentials $v_i$ of Eq. (3.2) can be decomposed as :
\begin{equation}
v_i\ =\ V_i\ +\ wU_i\ +\ w_{12}T_i\ \ \ \ \ (i=1,\ldots ,4)\ ,
\end{equation}
where $V_i,\ U_i$ and $T_i$ are functions of $x^{T2}$ and $P_L$.\par
For the combination of potentials we introduced in Eq. (2.20), we have to
reexpress the transverse matrices $\gamma ^T$ according to Eqs. (3.4). By
comparing Eq. (2.20) with Eqs. (3.2) and (3.6) we identify the indices
used in Eq. (2.20).\par
In order to project Eqs. (2.21) onto the subspaces of the components $\psi _i$
[Eq. (3.1)], it is useful to introduce projection matrices for these
subspaces. These are :
\widetext
\begin{eqnarray}
{\cal P}_1\ &=&\ \frac {1}{4} (1 + \gamma _{1L} \gamma _{2L})\
(1 + \gamma _{15} \gamma _{25})\ ,\ \ \ {\cal P}_2\ =\ \frac {1}{4}
(1 + \gamma _{1L} \gamma _{2L})\ (1 - \gamma _{15} \gamma _{25})\ ,\nonumber
\\
{\cal P}_3\ &=&\ \frac {1}{4} (1 - \gamma _{1L} \gamma _{2L})\
(1 + \gamma _{15} \gamma _{25})\ ,\ \ \ {\cal P}_4\ =\ \frac {1}{4}
(1 - \gamma _{1L} \gamma _{2L})\ (1 - \gamma _{15} \gamma _{25})\ .\nonumber
\\
\
\end{eqnarray}
\narrowtext
They satisfy the relations :
\widetext
\begin{equation}
{\cal P}_i {\cal P}_j\ =\ \delta _{ij} {\cal P}_j \ ,\ \ \
{\cal P}_i \Gamma _j\ =\
\delta _{ij} \Gamma _j \ \ \ \ (i,j=1,\ldots ,4)\ .
\end{equation}
\narrowtext
(The $\Gamma $'s are defined in Eq. (3.1).)\par
The general potential $V$ [Eq. (3.2)] can be expressed in terms of the
projection matrices ${\cal P}$ :
\begin{eqnarray}
V\ &=&\ \sum _{i=1}^4 a_i {\cal P}_i\ ,\nonumber \\
a_i\ &=&\ A_i\ +\ wB_i\ +\ w_{12}C_i\ ;
\end{eqnarray}
the relations between the sets $A_i,\ B_i,\ C_i$ and $V_i,\ U_i,\ T_i$
[Eq. (3.6)] are easily established. For the potential (2.20) we shall
consider below, we have the identifications :
\begin{eqnarray}
a_1\ &=&\ V_1\ +\ V_2\ +\ V_3\ +\ wU_4\ +\ w_{12}T_4\ ,\nonumber \\
a_2\ &=&\ V_1\ +\ V_2\ -\ V_3\ -\ wU_4\ -\ w_{12}T_4\ ,\nonumber \\
a_3\ &=&\ V_1\ -\ V_2\ +\ V_3\ -\ wU_4\ -\ w_{12}T_4\ ,\nonumber \\
a_4\ &=&\ V_1\ -\ V_2\ -\ V_3\ +\ wU_4\ +\ w_{12}T_4\ .
\end{eqnarray}
\par
The projectors (3.7) are mainly needed for the evaluation of the
exponential functions of $V$, present in Eqs. (2.21). One has the simple
property :
\begin{equation}
\exp \big (\sum _{i=1}^4 a_i {\cal P}_i \big )\ =\
\sum _{i=1}^4 {\cal P}_i\ e^{\displaystyle a_i}\ .
\end{equation}
\par
Equations (2.21) then yield the eight coupled (compatible) equations for
the four components $\psi _i \ (i=1,\ldots ,4)$ :
\widetext
\begin{eqnarray}
P_L\ e^{\displaystyle -a_3}\ \psi _3\ &-&\ M\ e^{\displaystyle a_4}\ \psi _4\
+\ \frac {2}{\hbar P_L}(W_{1S} - W_{2S}).p\ e^{\displaystyle -a_1}\ \psi _1\
=\ 0\ ,\nonumber \\
P_L\ e^{\displaystyle -a_4}\ \psi _4\ &-&\ M\ e^{\displaystyle a_3}\ \psi _3\
-\ \frac {2}{\hbar P_L}W_S.p\ e^{\displaystyle -a_2}\ \psi _2\ =\ 0\ ,
\nonumber \\
\frac {(m_1^2 - m_2^2)}{P_L}\ e^{\displaystyle -a_1}\ \psi _1\ &-&\
M\ e^{\displaystyle a_2}\ \psi _2\
+\ \frac {2}{\hbar P_L}W_S.p\ e^{\displaystyle -a_3}\ \psi _3\ =\ 0\ ,
\nonumber \\
\frac {(m_1^2 - m_2^2)}{P_L}\ e^{\displaystyle -a_2}\ \psi _2\ &-&\
M\ e^{\displaystyle a_1}\ \psi _1\
-\ \frac {2}{\hbar P_L}(W_{1S} - W_{2S}).p\ e^{\displaystyle -a_4}\ \psi _4\ =
\ 0\ ,\nonumber \\
\frac {(m_1^2 - m_2^2)}{P_L}\ e^{\displaystyle a_3}\ \psi _3\ &-&\
\frac {(m_1^2 - m_2^2)}{M}\ e^{\displaystyle -a_4}\
\psi _4\ +\ \frac {2}{\hbar P_L}W_S.p\ e^{\displaystyle a_1}\ \psi _1\ =\ 0\ ,
\nonumber \\
\frac {(m_1^2 - m_2^2)}{P_L}\ e^{\displaystyle a_4}\ \psi _4\ &-&\
\frac {(m_1^2 - m_2^2)}{M}\ e^{\displaystyle -a_3}\
\psi _3\ -\ \frac {2}{\hbar P_L}(W_{1S} - W_{2S}).p\ e^{\displaystyle a_2}\
\psi _2\ =\ 0\ ,\nonumber \\
P_L\ e^{\displaystyle a_1}\ \psi _1\ &-&\
\frac {(m_1^2 - m_2^2)}{M}\ e^{\displaystyle -a_2}\ \psi _2\ +\
\frac {2}{\hbar P_L}(W_{1S} - W_{2_S}).p\ e^{\displaystyle a_3}\ \psi _3\ =
\ 0\ ,\nonumber \\
P_L\ e^{\displaystyle a_2}\ \psi _2\ &-&\
\frac {(m_1^2 - m_2^2)}{M}\ e^{\displaystyle -a_1}\ \psi _1\ -\
\frac {2}{\hbar P_L}W_S.p\ e^{\displaystyle a_4}\ \psi _4\ =\ 0\ .\nonumber \\
\
\end{eqnarray}
\narrowtext
\par
By using the property $(W_S.p)(W_{1S} - W_{2S}).p = 0$, one easily deduces
from Eqs. (3.12) the following four equations :
\widetext
\begin{eqnarray}
W_S.p\ (P_L\ e^{\displaystyle -a_3}\ \psi _3)\ &=&\
W_S.p\ (M\ e^{\displaystyle a_4}\ \psi _4)\ ,\nonumber \\
(W_{1S} - W_{2S}).p\ (P_L\ e^{\displaystyle -a_4}\ \psi _4)\ &=&\
(W_{1S} - W_{2S}).p\ (M\ e^{\displaystyle a_3}\ \psi _3)\ ,\nonumber \\
(W_{1S} - W_{2S}).p\ (P_L\ e^{\displaystyle a_2}\ \psi _2)\ &=&\
(W_{1S} - W_{2S}).p\ \big ( \frac {(m_1^2 - m_2^2)}{M}\
e^{\displaystyle -a_1}\ \psi _1\big )
\ ,\nonumber \\
W_S.p\ (P_L\ e^{\displaystyle a_1}\ \psi _1)\ &=&\
W_S.p\ \big ( \frac {(m_1^2 - m_2^2)}{M}\
e^{\displaystyle -a_2}\ \psi _2\big )\ ,\nonumber \\
\
\end{eqnarray}
\narrowtext
which are useful during intermediate stages of calculation and for
consistency checks.\par
Among the four components $\psi _i$, it is the combination $\psi _3 + \psi _4$
which is the dominant one in the nonrelativistic limit. The quantum numbers
of the state are therefore determined by $\psi _3$ and $\psi _4$. For this
reason one must eliminate from Eqs. (3.12) $\psi _1$ and $\psi _2$ in terms
of $\psi _3$ and $\psi _4$ and finally keep $\psi _3$ or $\psi _4$. We shall
keep the component $\psi _3$ and express the eigenvalue equation in terms
of it; the calculations with $\psi _4$ parallel those with $\psi _3$ and
we shall indicate at the end the substitutions to use to obtain the
eigenvalue equation with $\psi _4$.\par
The third and seventh of Eqs. (3.12) allow us to express $\psi _1$ and
$\psi _2$ in terms of $\psi _3$ :
\begin{mathletters}
\begin{eqnarray}
\psi _1\ =\ e^{\displaystyle (a_1 - 2h)}\ \frac {2}{\hbar P^2}\
\big \{\ &-& e^{\displaystyle -2a_1}\ (W_{1S} - W_{2S}).p\
e^{\displaystyle a_3}\ \psi _3 \nonumber \\
&+& \frac {(m_1^2 - m_2^2)}{M^2}\ e^{\displaystyle -2(a_1 + a_2)}\
W_S.p\ e^{\displaystyle -a_3}\ \psi _3\ \big \}\ ,\\
\psi _2\ =\ e^{\displaystyle (a_2 - 2h)}\ \frac {2}{\hbar M P_L}\
\big \{\ &+& e^{\displaystyle -2a_2}\ W_S.p\
e^{\displaystyle -a_3}\ \psi _3 \nonumber \\
&-& \frac {(m_1^2 - m_2^2)}{P^2}\ e^{\displaystyle -2(a_1 + a_2)}\
(W_{1S} - W_{2S}).p\ e^{\displaystyle a_3}\ \psi _3\ \big \}\ ,
\end{eqnarray}
\end{mathletters}
where we have defined :
\begin{mathletters}
\begin{eqnarray}
e^{\displaystyle 2h}\ &=&\ 1\ -\ \frac {(m_1^2 - m_2^2)^2}{M^2 P^2}\
e^{\displaystyle -2(a_1 + a_2)} \\
&=&\ 1\ -\ \frac {(m_1^2 - m_2^2)^2}{M^2 P^2}\
e^{\displaystyle -4(V_1 + V_2)}\ ,
\end{eqnarray}
\end{mathletters}
the second equation, (3.15b), being valid only for the case of
potential (2.20).\par
One then obtains two independent equations for $\psi _3$ and $\psi _4$ :
\widetext
\begin{mathletters}
\begin{eqnarray}
M P_L\ e^{\displaystyle -a_4}\ \psi _4\ =\ &M^2& e^{\displaystyle a_3}\
\psi _3\nonumber \\
&+&\ \big (\frac {2}{\hbar P_L}\big )^2 W_S.p\ e^{\displaystyle -2h}\ \big \{\
e^{\displaystyle -2a_2}\ W_S.p\ e^{\displaystyle -a_3}\ \psi _3
\nonumber \\
&-&\ \frac {(m_1^2 - m_2^2)}{P^2}\ e^{\displaystyle -2(a_1 + a_2)}\
(W_{1S} - W_{2S}).p\ e^{\displaystyle a_3}\ \psi _3\ \big \}\ ,\\
\ \nonumber \\
M P_L\ e^{\displaystyle a_4}\ \psi _4\ =\ &P^2& e^{\displaystyle -a_3}\
\psi _3\nonumber \\
&+&\ \big (\frac {2}{\hbar P_L}\big )^2 (W_{1S} - W_{2S}).p\
e^{\displaystyle -2h}\
\big \{\ - e^{\displaystyle -2a_1}\ (W_{1S} - W_{2S}).p\
e^{\displaystyle a_3}\ \psi _3 \nonumber \\
&+&\ \frac {(m_1^2 - m_2^2)}{M^2}\ e^{\displaystyle -2(a_1 + a_2)}\
W_S.p\ e^{\displaystyle -a_3}\ \psi _3\ \big \}\ .
\end{eqnarray}
\end{mathletters}
\narrowtext
\par
Finally, eliminating $\psi _4$ through Eqs. (3.16) one obtains the
eigenvalue equation with $\psi _3$:
\widetext
\begin{eqnarray}
P^2 e^{\displaystyle -(a_3 + a_4)}\ &\psi _3&\ -\
M^2 e^{\displaystyle (a_3 + a_4)}\ \psi _3\nonumber \\
-\big (\frac {2}{\hbar P_L})^2 &e^{\displaystyle a_4}&\ W_S.p\
e^{\displaystyle -2h}\ \big \{\ e^{\displaystyle -2a_2}\ W_S.p\
e^{\displaystyle -a_3} \nonumber \\
&-& \frac {(m_1^2 - m_2^2)}{P^2}\ e^{\displaystyle -2(a_1 + a_2)}\
(W_{1S} - W_{2S}).p\ e^{\displaystyle a_3}\ \big \}\ \psi _3 \nonumber \\
-\big (\frac {2}{\hbar P_L}\big )^2 &e^{\displaystyle -a_4}&\
(W_{1S} - W_{2S}).p\ e^{\displaystyle -2h}\ \big \{\ e^{\displaystyle -2a_1}
(W_{1S} - W_{2S}).p\ e^{\displaystyle a_3} \nonumber \\
&-& \frac {(m_2^2 - m_2^2)}{M^2}\ e^{\displaystyle -2(a_1 + a_2)}
\ W_S.p\ e^{\displaystyle -a_3}\ \big \}\ \psi _3\ =\ 0\ .
\end{eqnarray}
\narrowtext
\par
Had we expressed $\psi _1$ and $\psi _2$ in terms of $\psi _4$ and
eliminated $\psi _3$ in terms of $\psi _4$, we would have obtained the
eigenvalue equation with $\psi _4$, which can be read off Eq. (3.17) with
the following substitutions {\it in the terms containing the spin
operators} : $a_3 \leftrightarrow -a_4$ and \\ $(m_1^2 - m_2^2)/P^2
\leftrightarrow (m_1^2 - m_2^2)/M^2$.\par
Equation (3.17) is a second order differential equation and has the form of
a generalized Pauli-Schr\"odinger equation. It is possible to bring it
into a more standard form by moving the momentum operators to the utmost
right and having recourse to wave function transformations. We shall
undertake these operations for the case when potential $V$ takes the form
(2.20) (or (3.10)) (scalar, pseudoscalar and vector potentials), the
calculations, as well as the final expressions, being relatively simple.
For simplicity of notation and a more transparent reading of the formulas,
we shall express the resulting equation in the c.m. frame. We use the
notations :
\widetext
\begin{equation}
{\bf S}\ =\ {\bf s_1 + s_2}\ ,\ \ \ {\bf L}\ =\ {\bf x}\times {\bf p}\ ,
\ \ \ {\bf J}\ =\ {\bf L} + {\bf S}\ ,\ \ \
S_{12}\ =\ \frac {4}{\hbar ^2} {\bf \frac {(s_1.x) (s_2.x)}{x^2}}\ .
\end{equation}
\narrowtext
\par
Upon bringing, in Eq. (3.17) and for the case of potential (2.20) (or (3.10)),
the momenntum operators to the utmost right, one finds that the quadratic
terms in ${\bf p}$ sum up to yield, apart from a multiplicative potential
dependent factor, the operator ${\bf p^2}$. However, one also meets terms
proportional to $i\hbar {\bf x.p}$. To get rid of the latter terms, it is
necessary to use the change of function :
\begin{equation}
\psi _3\ =\ e^{\displaystyle (2U_4 + 2T_4 + h)}\ \big \{\ \frac {1}{2}
(1+S_{12})\ e^{\displaystyle 2V_1}\ +\ \frac {1}{2} (1-S_{12})\
e^{\displaystyle 2V_2}\ \big \}\ \phi _3\ .
\end{equation}
\par
After some lengthy calculations (see Appendix B for useful formulas that
are utilized for the spin and orbital angular momentum operators) one ends up
with the following eigenvalue equation :
\pagebreak
\widetext
\begin{eqnarray}
\bigg \{ &\ &e^{\displaystyle 4(U_4 + T_4)}\ \bigg [ \ \frac {P^2}{4}
e^{\displaystyle 4V_2}\ -\ \frac {M^2}{4} e^{\displaystyle 4V_1}\ -\
\frac {(m_1^2 - m_2^2)^2}{4M^2}\ e^{\displaystyle -4V_1}\ +\
\frac {(m_1^2 - m_2^2)^2}{4P^2}\ e^{\displaystyle -4V_2}\ \bigg ]\ \
\nonumber \\
&-&\ {\bf p^2}\ -\
4\hbar ^2 {\bf x^2}\ (\ V_1'+ V_2'+V_3'-U_4'+T_4'+h'\ )^2\nonumber \\
&+&\ 6\hbar ^2\ (\ V_1'+V_2'+V_3'-U_4'+T_4'+h'\ )\
+\ 4\hbar ^2 {\bf x^2}\ (\ V_1''+V_2''+V_3''-U_4''+T_4''+h''\ )\nonumber \\
&-&\ {\bf L^2} \frac {1}{{\bf x^2}} (e^{\displaystyle 4T_4} - 1)\nonumber \\
&-&\ 4{\bf S^2}\  \bigg [\ (\ V_1'+V_2'+V_3'+U_4'+T_4'+h'\ )\
(\ 1+4{\bf x^2}U_4'\ )\nonumber \\
&\ &\ \ \ \ \ \ -\  (\ U_4'+2{\bf x^2}U_4''\ )\ -\
\frac {1}{4{\bf x^2}} (e^{\displaystyle 4T_4} - 1)\ \bigg ]\nonumber \\
&+&\ 4{\bf \frac {(S.x)^2}{x^2}}\ \bigg [\ (\ V_1'+V_2'+2U_4'+T_4'+h'\ )\
\big (\ 1+4{\bf x^2}(V_3'+U_4')\ \big )\nonumber \\
&\ &\ \ \ \ \ \ -\ 2{\bf x^2}(V_3''+U_4'')
\ -\ \frac {1}{2{\bf x^2}} (e^{\displaystyle 4T_4} - 1)\ \bigg ]\nonumber \\
&+&\ 4\big (\ {\bf L.S + S^2 - \frac {(S.x)^2}{x^2}}\ \big )\
e^{\displaystyle 2T_4}\  \bigg [\ \frac {1}{{\bf x^2}} \sinh ^2(V_1-V_2)
\ -\ \frac {1}{2{\bf x^2}} (e^{\displaystyle 2T_4} - 1)\ \cosh 2(V_1-V_2)\
\nonumber \\
&\ &\ \ \ \ \ \ +\ (\ V_1'+V_2'+2U_4'+h'\ )\ \cosh 2(V_1-V_2) \ -\
(V_1'-V_2') \sinh 2(V_1-V_2)\ \bigg ]\nonumber \\
&+&\ 4 \big [\ \frac {{\bf (S.x)^2}}{\hbar ^2 {\bf x^2}}\ ,\ {\bf L.S}\ \big ]
\ e^{\displaystyle 2T_4}\ \bigg [\ \frac {1}{2{\bf x^2}}
(e^{\displaystyle 2T_4} - 2)\ \sinh 2(V_1-V_2)\nonumber \\
&\ &\ \ \ \ \ \ -\ (\ V_1'+V_2'+2U_4'+h'\ )\ \sinh 2(V_1-V_2)\ +\
(V_1'-V_2')\ \cosh 2(V_1-V_2)\ \bigg ]\nonumber \\
&-&\ 4(m_1^2 - m_2^2)\ (\frac {1}{P^2} + \frac {1}{M^2})\ {\bf L.(s_1-s_2)}\
e^{\displaystyle 2[T_4 - (V_1+V_2) - h]}\ (V_1'+V_2')\nonumber \\
&-&\ 4(m_1^2 - m_2^2)\ (\frac {1}{P^2} - \frac {1}{M^2})\
\big [\ \frac {{\bf (S.x)^2}}{\hbar ^2 {\bf x^2}}\ ,\ {\bf L.(s_1-s_2)}\
\big ]\
e^{\displaystyle 2[T_4 - (V_1+V_2) - h]}\ (V_1'+V_2')\ \bigg \} \
\phi _3  \nonumber \\
&\ &\ \ \ \ \ \ \ \ \ \ \ \ \ \ \ \ \ \ \ \ \ \ \ \ \ \ \ \ \ \ \ \ \ \
\ \ \ \ \ \ \ \ \ \ \ \ \ \ \ \ \ \ \ \ \ \ \ \ \ \ \ \ \ \ \ \ \ \ \ \
\ \ \ \ \ \ \ \ \ \ \ \ \ \ \ \ =\ 0\ .\nonumber \\
\
\end{eqnarray}
\narrowtext
\par
We recall that the functions $V_1,\ V_2,\ V_3,\ U_4$ and $T_4$ correspond
to the scalar ($V_1$), timelike vector ($V_2$), pseudoscalar ($V_3$) and
spacelike vector ($U_4$ and $T_4$) potentials. The function $h$ is defined
in Eq. (3.15). We have also defined :
\begin{equation}
F'\ \equiv \ \frac {\partial F}{\partial {\bf x^2}}\ ,\ \ \
F''\ \equiv \ \frac {\partial ^2 F}{(\partial {\bf x^2})^2}\ .
\end{equation}
\par
Equation (3.20) is the main result of the present paper. It corresponds to
the reduction of the initial covariant wave equations (2.1), by means of
the parametrizations (2.17) and (2.20) and of the wave function
transformations and decomposition (2.18), (3.1), (3.19), to the
Pauli-Schr\"odinger form, in which $\phi _3$ is a $2\times 2$ matrix wave
function.\par
The above reduction can be realized with respect to the component $\psi _4$
as well. In this case the wave function transformation (3.19) is
replaced by the following one :
\begin{equation}
\psi _4\ =\ e^{\displaystyle (2U_4 + 2T_4 + h)}\ \big \{\ \frac {1}{2}
(1+S_{12})\ e^{\displaystyle 2V_2}\ +\ \frac {1}{2} (1-S_{12})
\ e^{\displaystyle 2V_1}\ \big \}\ \phi _4\ .
\end{equation}
\par
The equation satisfied by $\phi _4$ is the same as Eq. (3.20), except a
global change of sign in front of the terms proportional to the
commutators $[\ {\bf (S.x)^2}/(\hbar ^2 {\bf x^2})\ ,\ {\bf L.S}\ ]$
and $[\ {\bf (S.x)^2}/(\hbar ^2 {\bf x^2})\ ,\ {\bf L.(s_1-s_2)}\ ]$ .\par
In order to reconstitute the whole wave function $\psi$ from $\psi _3$,
it is also necessary to calculate $\psi _4$ from $\psi _3$ (or conversely,
$\psi _3$ from $\psi _4$ if the latter is first obtained). To this end,
the first two of Eqs. (3.13) kinematically yield the desired relations in
the sectors of states corresponding to the quantum numbers $j=\ell \ $.
However, in the sectors with quantum numbers $j\ne \ell $ these
equations do not allow one to obtain kinematically the relationship
between $\psi _3$ and $\psi _4\ $. It is therefore necessary to go back
to the two coupled equations (3.16). Either of these equations can be used
to express $\psi _4$ in terms of $\psi _3\ $. As they stand, these
equations involve second order differential operators. In order to obtain
a relationship involving at most first order differential operators, one
eliminates the operator ${\bf p^2}$ (see for this Eq. (B8)) by the use of
Eq. (3.17). The relation giving $\psi _4$ in terms of $\psi _3$ is
presented in Appendix B [Eq. (B9)] where we also explain how to express
$\psi _3$ in terms of $\psi _4\ $.\par
For completeness, we rewrite, in Appendix B [Eqs. (B10)], Eqs. (3.14)
giving $\psi _1$ and $\psi _2$ in terms of $\psi _3$, in more explicit
forms when potential $V$ has the form (2.20).\par
Equation (3.20) can be solved by decomposing the wave function $\phi _3$
along radial wave functions and the generalized spherical harmonics,
${\cal Y}_{j\ell s}^m\ $, eigenfunctions of the operators ${\bf J^2, J_z,
L^2, S^2}\ $.\par
When all potentials are present, and in the unequal mass case, there are only
two sectors of quantum numbers where $\phi _3$ is an eigenfunction of the
latter operators : i) $s=0,\ \ell =0,\ j=0\ $; ii) $s=1,\ \ell =1,\ j=0\ \ $.
For the other sectors $\phi _3$ is a combination of different eigenfunctions :
$s=0$ and $s=1$ for the sectors with $\ell =j\ $; $\ell = \vert j \pm 1\vert$
for the sectors with $s=1\ $. In the equal mass case, a simplification occurs,
in that for the sectors with $\ell =j$, the subsectors with $s=0$ and $s=1$
disantangle. A similar simplification also occurs in the unequal mass case
when $V_1$ and $V_2$ are absent, or when $V_1+V_2=0$ (or constant).
Complete diagonalization occurs only for harmonic oscillators.\par
The presence in Eq. (3.20) of the commutators $[\ {\bf (S.x)^2}/(\hbar ^2
{\bf x^2})\ ,\ {\bf L.S}\ ]$ and $[\ {\bf (S.x)^2}/(\hbar ^2 {\bf x^2})\ ,\
{\bf L.(s_1 - s_2)}\ ]$, might suggest that some of the energy eigenvalues
are not real. These operators connect only off-diagonal matrix elements of
the sectors $\ell =\vert j\pm 1\vert\ $. As is known, such matrix elements
contribute only with the modulus squared of their values, and therefore the
resulting energy eigenvalues are real. This also explains why, on solving
either of
Eq. (3.20) or the equivalent equation of $\phi _4\ $, where the signs in
front of the above commutators are changed, one obtains the same energy
eigenvalues.\par
In this connection, it is worth emphasizing that if, instead of considering
the complete norm (2.19), where in any event the hermiticity property of the
energy operator is manifest provided $V$ is superficially hermitian, one
tries to construct the scalar product or the norm for $\phi _3$ directly
from Eq. (3.20), one has to consider as the adjoint of
$\phi _3$ the function $\phi _4^{\dagger }$ rather than $\phi _3^{\dagger }\ $,
precisely because of the antihermitian nature of the abovementioned
commutators. If Eq. (3.20) is written in compact form as $L_3 \phi _3 =0\ $,
then following standard methods \cite{s1,f,lcl}, one ends up with a norm
proportional to $\int \ d^3{\bf x}\ Tr\ \phi _4^{\dagger } \
\frac {\partial L_3}{\partial P^2}\ \phi _3\ $, the hermitian conjugate of
which is nothing but $\int \ d^3{\bf x}\ Tr\ \phi _3^{\dagger }\
\frac {\partial L_4}{\partial P^2}\ \phi _4\ $, because $L_3^{\dagger } =
L_4\ $, where $L_4$ is the wave equation operator of $\phi _4\ $, satisfying
$L_4 \phi _4 = 0\ $.\par
Finally, we observe that when the potentials $V_1$ and $V_2$ are absent, then
$\phi _3$ and $\phi _4$ satisfy the same eigenvalue equation, and yet they
are not identical in general (see Eq. (B9)). This means that the energy
spectrum has then degeneracies coming from the sectors with
$\ell = \vert j \pm 1\vert \ $. These degeneracies increase in the case of
potentials yielding harmonic oscillators.\par
\pagebreak

\section{Infinite mass limit}

In this section we shall study the limit of Eq. (3.20)
when one of the particles, 2, say, becomes infinitely
massive. It is known that the Bethe-Salpeter equation
does not have a simple behavior in this limit.
The origin of this difficulty lies in the fact that
the kernel of the Bethe-Salpeter equation contains
only irreducible (crossed ladder) diagrams, which do not
have suitable limits. In order to reproduce in this limit,
from the Bethe-Salpeter equation, the
one-particle Dirac equation in the presence of the potential
created by the infinitely massive particle, one must take into
account contributions coming from the infinite number
of crossed ladder diagrams \cite{b,ng}.\par

The situation is much simpler for the wave equations of
Constraint Theory. Here, the potential is calculated from the diagrams
of the off-mass shell scattering amplitude, to which one adds
``constraint diagrams'', obtained from reducible diagrams with
the use of constraint (2.4) [Eq. (2.10)]. To a given formal order
of perturbation theory, the sum of all ladder and crossed ladder type
Feynman diagrams (without vertex corrections) does have a simple limit :
the sum of the $n^{{\rm th}}$ order diagrams (in coupling
constant squared) contains the factor $\delta (k_{10})\cdots \delta (k_{n-1
0})$, where the $k$'s are the internal momenta of the loops ($n\geq 2$)
\cite{b,ng}.
This term is then cancelled by the dominant term of the
``constraint diagrams'' of ${n^{{\rm th}}}$ order \cite{s2} (see Appendix C).
Therefore, in the limit $m_{2} \rightarrow \infty$ it is only
the one-particle exchange diagram (including vacuum polarization)
which survives and leads to the one-particle Dirac equation
with the corresponding static potential.\par

The expressions of the potentials calculated from such diagrams
are presented in Appendix A. They essentially have the
energy dependence of the form
\begin{equation}
V\ =\ \frac {1} {2P_{L}}\ V^{(0)} \ ,
\end{equation}
and, therefore, when $m_2$ goes to infinity their behavior is :
\begin{equation}
\lim _{m_2 \rightarrow \infty} V\ =\
\frac {1} {2m_2} V^{(0)}\ +O\bigl
(\frac {1} {m_{2}^{2}}\bigr ) \ .
\end{equation}
\par
The contributions coming from higher order diagrams
are of order $O(1/m_{2}^{2})$. If one replaces in Eq. (2.23)
the various potentials by their leading behaviors (4.2), one
immediately finds that Eq. (2.23) reduces to the Dirac equation
of particle 1 in the presence of the external static
scalar and timelike vector potentials $V_{1}^{(0)}$ and $V_{2}^{(0)}$.
It is, however, also instructive to study the above
limit in Eq. (3.20), where the contributions of each potential
are more involved.\par

To this end we expand each potential
$V_{a}\  (a=1,\ldots ,4,\ V_{4}$ representing here either
$U_{4}$ or $T_{4})$ up to its next to the leading
term:
\begin{equation}
V_{a}\ \simeq \ {\frac {1} {2m_{2}}}V_{a}^{(0)}\ +\
{\frac {1} {m_{2}^{2}}}\ W_{a}^{(1)} \ .
\end{equation}
\par
The total energy $P_{L}$ is decomposed as:
\begin{equation}
P_{L}\ =\ m_{2}\ +\ E \ .
\end{equation}
\par
The terms inside the brackets containing $V_{1}$ and $V_{2}$ have
the following limit:
\widetext
\begin{eqnarray}
\lim _{m_2 \rightarrow \infty}
\bigg [\ { {\frac {P^{2}} {4}}\ e^{\displaystyle 4V_{2}}- {\frac {M^{2}} {4}}\
e^{\displaystyle 4V_{1}}-{\frac {(m_{1}^2-m_{2}^2)^{2}} {4M^2}}\
e^{\displaystyle -4V_{1}}
+{\frac {(m_{1}^{2}-m_{2}^{2})^{2}} {4P^{2}}}\
e^{\displaystyle -4V_{2}}}\ \bigg ]\ =\nonumber \\
\bigg [\ \bigl ( E + V_2^{(0)} \bigr )^2\ -\
\bigl ( m_1 + V_1^{(0)} \bigr )^2\ \bigg ]\ .
\end{eqnarray}
\narrowtext
Notice that the terms $W_{a}^{(1)}$ have disappeared, although
the brackets were containing terms of
the order of $m_{2}^{2}$. It is also evident that the
factor $\exp \big (4(U_{4}+T_{4})\big )$ in front of
the brackets in Eq. (3.20) will contribute as unity.
\par
In the remaining terms of Eq. (3.20), besides the kinetic energy term,
it is only those containing $h',\ h''$ and $\exp (-2h)$
which will survive. They have the limits:
\begin{eqnarray}
\lim _{m_2 \rightarrow \infty } \frac {2}{m_2}\ e^{\displaystyle -2h}\ &=&\
\bigl [\ (E + V_2^{(0)})\ +\ (m_1 + V_1^{(0)})\ \bigr ]^{-1} \ ,
\nonumber \\
\lim _{m_2 \rightarrow \infty } h'\ &=&\
{\frac {\bigl (V_{2}^{(0)}{}'+V_{1}^{(0)}{}'\bigr )}
{2\ \bigl [\ (E+V_{2}^{(0)})\ +\ (m_1+V_{1}^{(0)})\ \bigr ]}}\ .
\end{eqnarray}
\par
After averaging over the spin states of particle $2$, one
ends up with the following
equation (where $\varphi$ designates the limit of $\phi_{3}/m_{2}$):
\begin{eqnarray}
\bigg \{\ (E&+&V_{2}^{(0)})^{2}\ -\ {\bf p^2}\ -\ (m_{1}+V_{1}^{(0)})^{2}
\nonumber \\
&-&4\hbar^{2}{\bf x^2}h'{}^{2}+6\hbar^{2}h'
+4\hbar^{2}{\bf x^2}h'' +8h'{\bf L.s_{1}}\ \bigg \}\varphi \ =\ 0 \ .
\end{eqnarray}
\par
This equation is nothing but the reduced form of the Dirac equation in
the presence of the static scalar and timelike vector potentials
after the decomposition $\psi =\biggl ({^{\psi _{+}}_{\psi _{-}}}\biggr )$
and the change of function
$
\psi_{+}=\bigl (E+V_{2}^{(0)}+m_1+V_{1}^{(0)})^{1/2}\varphi
$
are used (in the basis where $\gamma _0$ is diagonal).
The latter is the limit (up to a multiplicative constant
factor proportional to $m_{2}^{-1/2}$),
when $m_{2}$ goes to infinity, of the change of function
(3.19).\par
\pagebreak

\section{Nonrelativistic limit to order ${\bf 1/c^2}$}

We study, in this section, the nonrelativistic
limit of Eq. (3.20) to order $1/c^{2}$,
the gauge transformation property of the resulting
Hamiltonian for the electromagnetic interaction case and the structure
of confining scalar potentials.\par

\subsection{General case in perturbation theory}

To recover the explicit $c$ dependence of the various terms in Eq.
(3.20) (where $c$ had been set equal to 1) we must  adjoin
a $c$ factor to each of the terms $\hbar ,\  {\bf p,\ s_{1},\
s_{2},\  L}$ and a $c^{2}$ factor to $m_{1}$ and $m_{2}$.
On the other hand, at leading order, the
potentials have, in perturbation theory, the dependence on
$P_{L}$ of the type $1/(2P_L)$ (see Appendix A).
Therefore, we expand the potentials as follows :
\begin{equation}
V_{a}\ =\ \frac {1}{2P_{L}}\ V_a^{(0)}\ +\
\frac {1}{2P_{L}^{2}}\ V_a^{(1)}\
+\ O(P_{L}^{-3}) \ \ \ \ (a=1,\ldots ,4) \ .
\end{equation}
Here, $V_{a}^{(0)}/(2P_L)$ is the contribution of the one-particle exchange
diagram (or an effective representation of it) and $V_{a}^{(1)}/(2P_L^2)$ is
the leading contribution coming from the higher order diagrams.
(In the perturbation theory of the present formalism, it is the
two-particle exchange diagrams which provide
the latter term.)\par

To carry out the expansion, we split the total
energy into the total mass and the binding energy $E$ :
\begin{equation}
P_{L}\ =\ M\ +\ E \ .
\end{equation}
\par
The term $E$, which is an eigenvalue, is brought to
the utmost right and replaced there by the Hamiltonian $H$.
In expressions containing the term $H^{2}$, which enters only in
$O(c^{-2})$ quantities, one replaces $H$ by its nonrelativistic
expression.\par
\pagebreak
The resulting Hamiltonian has the the following expression
(in the c.m. frame) :
\widetext
\begin{eqnarray}
H\ =&\ &\frac {\bf p^2} {2 \mu}\ +\ V_{1}^{(0)}\ -\ V_{2}^{(0)}\
-\ {\frac {1} {8}}\ \bigl (\ {\frac {1} {m_{1}^{3}}}
+ {\frac {1} {m_{2}^{3}}}\ \bigr )\ ({\bf p^2})^{2}\nonumber \\
&-& {\frac {1} {2m_{1}m_{2}}}\
\bigl [\ {\bf p^2}\ ,\ U_{4}^{(0)}+T_{4}^{(0)}\ \bigr ]_+\
-\ {\frac {(m_{1}^{2}+m_{2}^{2})} {4m_{1}^{2}m_{2}^{2}}}\
\bigl [\ {\bf p^2}\ ,\ V_{1}^{(0)}\ \bigr ]_{+}\nonumber \\
&-& {\frac {1}{2M}}\ (\ V_{1}^{(0)}-V_{2}^{(0)}\ )^{2}\ +\
{\frac {1}{M}}\ (\ V_{1}^{(1)}-V_{2}^{(1)}\ ) \nonumber \\
&-& {\frac {{\hbar}^{2}} {4m_{1}m_{2}}}\ \Delta
\biggl [\ {V_{1}^{(0)}+V_{2}^{(0)}+V_{3}^{(0)}-U_{4}^{(0)}
+T_{4}^{(0)}+{\frac {(m_{1}-m_{2})^{2}}
{2m_{1}m_{2}}}(V_{1}^{(0)}+V_{2}^{(0)})}\ \biggr ] \nonumber \\
&+& {\frac {1} {m_{1}m_{2}}}\ {\bf L^2}\ {\frac {T_{4}^{(0)}} {{\bf x^2}}}\
+\ {\frac {1} {m_{1}m_{2}}}\ {\bf S}^{2}\ \left[\ {-{\frac {1} {2}}
\Delta U_{4}^{(0)}+V_{3}^{(0)}{}'+U_{4}^{(0)}{}'+T_{4}^{(0)}{}' }\ \right]
\nonumber \\
&-& {\frac {1} {m_{1}m_{2}}}\ {\bf \frac {(S. x)^{2}} {x^{2}} }\
\left [\ { -{\frac {1} {2}} \Delta (V_{3}^{(0)}+U_{4}^{(0)})
+3(V_{3}^{(0)}{}'+U_{4}^{(0)}{}')+T_{4}^{(0)}{}'-{\frac {T_{4}^{(0)}}
{{\bf x^2}}}}\ \right ] \nonumber \\
&-& {\frac {1} {m_{1}m_{2}}}\ {\bf L.S}\
\left [\ {V_{1}^{(0)}{}'+V_{2}^{(0)}{}'+2U_{4}^{(0)}{}'
-{\frac {T_{4}^{(0)}} {{\bf x}^{2}}}
+{\frac {(m_{1}-m_{2})^{2}}
{2m_{1}m_{2}}}(V_{1}^{(0)}{}'+V_{2}^{(0)}{}')}\ \right ] \nonumber \\
&-& {\frac {1} {2}}\ \bigl (\ { {\frac {1} {m_{1}^{2}}} -{\frac {1}
{m_{2}^{2}}} }\ \bigr )\ {\bf L.(s_{1} - s_{2})}\
\bigl (\ V_{1}^{(0)}{}'+V_{2}^{(0)}{}'\ \bigr ) \ .\nonumber \\
\
\end{eqnarray}
\narrowtext
Here, $\mu = m_{1}m_{2}/ {M}$, $\Delta V$ is
the Laplacian of $V$, $\Delta V=6V'+4{\bf x^{2}}V''$, and
the derivatives are with respect to ${\bf x^{2}}\ $. $[\ ,\ ]_{+}$
is the anticommutator. To obtain Eq. (5.3) we
also used the wave function transformation :
\widetext
\begin{eqnarray}
\phi_{3}\ =\ \left[\ { 1-{\frac {1} {M}}(U_{4}^{(0)}+T_{4}^{(0)})
-{\frac {1} {2M}}V_{1}^{(0)}-{\frac {(m_{1}^{2}+m_{2}^{2}-m_{1}m_{2})}
{2m_{1}m_{2}M}}V_{2}^{(0)}}\ \right]\ \phi \ ,\nonumber \\
\
\end{eqnarray}
\narrowtext
to bring the Hamiltonian into a manifestly hermitian form.
The term containing
$[\ {\bf (S.x)^2}/(\hbar ^2 {\bf x^2})\ ,\ {\bf L.S}\ ]$ in Eq.
(3.20), which is formally of order $c^{-2}$, does
not actually contribute, for it vanishes in diagonal
matrix elements of the eigenvalues of  the
nonrelativistic Hamiltonian.
\par
\pagebreak

\subsection{Electromagnetic interaction}

We now consider the case of the electromagnetic interaction.
It is known that bound state energies are gauge invariant \cite{by,ffh}.
We check this property to order $c^{-2}$. We set in
Eq. (5.3) $V_{1}=V_{3}=0$. Furthermore, considering
the photon propagator [Eqs. (A6) and (A7)] we deduce
that $U_{4}$ must be split into two terms, $V_{2}$ and
a pure gauge term $\widetilde U$; $T_{4}$ is then equal to
$2{\bf x^2}\widetilde U'$ :
\begin{equation}
U_{4}\ =\ V_{2}\ +\ \widetilde U\ ,\ \ \ \
T_{4}\ =\ 2{\bf x^2}\ \widetilde U'\ .
\end{equation}
\par
After this splitting is introduced, the Hamiltonian (5.3) for
the electromagnetic interaction takes the form :
\widetext
\begin{eqnarray}
H\ =&\ & {\frac {{\bf p^2}} {2\mu}}\ -\ V_{2}^{(0)}\ -\ {\frac {1} {8}}\
\bigl (\ { {\frac {1} {m_{1}^{3}}}+ {\frac {1} {m_{2}^{3}}} }\ \bigr )\
({\bf p^2})^2\nonumber \\
&-& {\frac {1} {m_{1}m_{2}}}\ (\ V_{2}^{(0)}+\widetilde U^{(0)}\ )\
{\bf p}^{2}\
-\ {\frac {2\widetilde U^{(0)}{}'} {m_{1}m_{2}}}\ x_{i}x_{j}p_{i}p_{j}
\nonumber \\
&+& {\frac {2i \hbar} {m_{1}m_{2}}}\ (\ V_{2}^{(0)}{}'
+2\widetilde U^{(0)}{}'\ )\
{\bf x.p}\ +\ {\frac {i \hbar} {m_{1}m_{2}}}\ \Delta \widetilde U^{(0)}\
{\bf x.p} \nonumber \\
&+& {\frac {\hbar ^{2}} {4m_{1}m_{2}}}\ \Delta \left[\ {
{\frac {(6m_{1}m_{2}-m_{1}^{2}-m_{2}^{2})} {2m_{1}m_{2}}}\ V_{2}^{(0)}
+3\widetilde U ^{(0)} }\ \right]\nonumber \\
&+& {\frac {\hbar ^{2}} {2m_{1}m_{2}}}\ \Delta ({\bf x^2}\widetilde
U^{(0)}{}')\
-\ {\frac {1} {2M}}\ (V_{2}^{(0)})^{2}\  -\ {\frac {1} {M}}V_{2}^{(1)}
\nonumber \\
&+& {\frac {1} {m_{1}m_{2}}}\ {\bf S^2}\ \bigl (\ {
-{\frac {1} {2}}\Delta V_{2}^{(0)}+V_{2}^{(0)}{}' }\ \bigr )\
-\ {\frac {1} {m_{1}m_{2}}}\  {\bf \frac {(S.x)^{2}} {x^{2}}}\
\bigl (\ { -{\frac {1} {2}}\Delta V_{2}^{(0)}+3V_{2}^{(0)}{}'}\ \bigr )
\nonumber \\
&-& {\frac {1} {2}}\ \bigl (\ { {\frac {1} {m_{1}^{2}}}+
{\frac {1} {m_{2}^{2}}}+ {\frac {4} {m_{1}m_{2}}}}\ \bigr )\
{\bf L.S}\ V_{2}^{(0)}{}' \
-\ {\frac {1} {2}}\ \bigl (\ { {\frac {1} {m_{1}^{2}}}-{\frac {1}
{m_{2}^{2}}}}\
\bigr )\ {\bf L.(s_{1} - s_{2})}\ V_{2}^{(0)}{}' \ .\nonumber \\
\
\end{eqnarray}
\narrowtext
\par
We notice that the gauge function $\widetilde U^{(0)}$ has
disappeared from the spin dependent terms. To prove gauge invariance
of the energy corresponding to the Hamiltonian (5.6), we must show
that a unitary wave function transformation removes the
arbitrary gauge function $\widetilde U^{(0)}$ from $H$ and brings it
into its form of the Feynman gauge (in which $\widetilde U^{(0)}=0$),
which is taken here as the reference gauge. \par

It can be checked that the following wave function
transformation
\begin{eqnarray}
\phi  &\simeq & (\ 1+iG\ )\ \phi_{F} \ , \nonumber \\
G &=& {\frac {1} {2 \hbar M}}\ (\ \widetilde U^{(0)}\ {\bf p.x}+
{\bf x.p}\ \widetilde U^{(0)}\ )\
\simeq \ -{\frac {2\mu} {4i \hbar ^{2}M}}\ \bigl [\ H\ ,\
\int^{{\bf x^2}} \widetilde U^{(0)}\ d{\bf x^2}\ \bigr ]\ ,
\end{eqnarray}
where $\phi_{F}$ is the wave function in the Feynman gauge, removes
from Eq. (5.6) the existing
$\widetilde U^{(0)}$ dependent terms. [Actually,
it can be shown, by using the connection of the Constraint
Theory wave function with the Bethe-Salpeter wave function \cite{s2}, that
transformation (5.7) is the three-dimensional reduction,
to the leading order of the present approximation, of the gauge
transformation of the Bethe-Salpeter wave-function \cite{lkzbcm}.]
\par
The transformation (5.7) makes appear, however, in Eq. (5.6),
a new $\widetilde U^{(0)}$ dependent term, coming from the
commutation of $G$ with $V_{2}^{(0)}$ and equal to
$2{\bf x^2}\widetilde U^{(0)} {{V_{2}^{(0)}{}'}/{M}}$.
This term must be compensated by the gauge dependence of the second
order term $V_{2}^{(1)}$, present in
Eq. (5.6). If we designate by $V_{2F}^{(1)}$ the expression of
$V_{2}^{(1)}$ in the Feynman gauge, then $V_{2}^{(1)}$ is related
to $V_{2F}$ by the relation:
\begin{equation}
V_{2}^{(1)}\ =\ V_{2F}^{(1)}\ +\ 2{\bf x^2}\ \widetilde U ^{(0)}\
V_{2}^{(0)}{}'\ .
\end{equation}
\par
The expression of $V_{2F}^{(1)}$ was calculated
from the two photon exchange diagrams by
Rizov, Todorov and Aneva \cite{rta} in the Quasipotential Theory for
spin- ${ {1}/{2} }$ - spin- $0$ particle systems. The perturbation
theory in the Quasipotential Theory is very similar to that of
Constraint Theory [Eq. (2.10)], except that in the latter
case off-mass shell scatternig amplitudes are considered and
nowhere extrapolations from on-shell
quantities are used. These differences do not, however, show up
at $O(\alpha^{4})$. Furthermore, $V_{2}$ is a spin independent quantity
to this order. One finds $V_{2F}^{(1)}=-(V_{2}^{(0)})^{2}$.
(We also checked this result from Eq. (2.10) in the
two-fermion case.) One
then obtains for $V_{2}^{(1)}$ the following expression:
\begin{equation}
V_{2}^{(1)}\ =\ -(V_{2}^{(0)})^{2}\ +\ 2{\bf x^2}\
\widetilde U^{(0)}\ V_{2}^{(0)}{}' \ .
\end{equation}

Therefore, the Hamiltonian (5.6), together with the expression
(5.9) of $V_{2}^{(1)}$, provides a gauge invariant wave equation to
order $c^{-2}$.
Its expression in the Feynman gauge is obtained by setting in Eq.
(5.6) $\widetilde U^{(0)}=0$.
\par
The Breit Hamiltonian is obtained by using the Coulomb gauge.
In the present three-dimensional formalism and in the c.m frame,
the expressions of the photon propagator in the Coulomb gauge and in the
Landau gauge
coincide in lowest order [Eq. (A10)].
The latter corresponds to the choice $\widetilde U^{(0)}=
-{V_{2}^{(0)}}/ {2}$. By replacing $\widetilde U^{(0)}$ with this
expression in Eqs. (5.6) and (5.9), one recovers
the Breit Hamiltonian \cite{bsblp}. (Notice that the term
$\Delta\widetilde U^{(0)}{\bf x.p}$ vanishes, for
$\Delta \widetilde U^{(0)}$ is now a $\delta$-function. For the same
reason the term $\Delta V_{2}^{(0)} {\bf (S.x)^{2}/x^2}$
can be replaced by $\Delta V_{2}^{(0)} {\bf S^2}/3\ $.)
In this gauge all quadratic terms in
$V_{2}^{(0)}$ (replaced by its Coulombic expression (A9) )
disappear. This explains, once more, why in the Coulomb gauge one
obtains the correct $O(\alpha^{4})$ effects with one
photon exchange diagrams only.\par
The reconstitution of a covariant electromagnetic potential from the
previous results is, however, model dependent and sensitive to implicit
assumptions made about the contributions of the multiphoton exchange
diagrams. A first model is provided by the Todorov potential \cite{t},
introduced in Quasipotential Theory and later used in spectroscopic
calculations (for electromagnetic as well as for short distance quark
interactions) by Crater, Van Alstine and collaborators \cite{cva1,cbwva}.
In our notations, it corresponds to the choice (in the Feynman gauge) :
\begin{eqnarray}
V_2\ &=&\ U_4\ =\ \frac{1}{4}\ln \bigl (\ 1 + \frac {2V_2^{(0)}}{P_L}\
\bigr )\ ,\ \ \ \ \ T_4\ =\ 0\ ,\nonumber \\
V_2^{(0)}\ &=&\ \frac {\alpha }{r}\ ,\ \ \ \ r\ =\ \sqrt {-x^{T2}}\ .
\end{eqnarray}
\par
It was shown in Ref. \onlinecite{bcs} that for large values of the
coupling constant ($\alpha \sim 1/2$) this potential leads to an instability
of the vacuum state and might play a role in mechanisms of spontaneous
chiral symmetry breaking.\par
A second potential was also considered in Ref. \onlinecite{bcs}; it
corresponds to the choice (in the Feynman gauge) :
\begin{equation}
V_2\ =\ U_4\ =\ \frac {1}{2}\frac {V_2^{(0)}}{(P_L + V_2^{(0)})}\ ,
\ \ \ \ \ T_4\ =\ 0\ ,
\end{equation}
$V_2^{(0)}$ being the same as in Eq. (5.10). This potential has the
property of being regular at the origin and does not lead to an
instability of the vacuum state.
\par

\subsection{Scalar potential}

We now turn to the case of scalar potentials and, for simplicity, shall
ignore the other potentials. We first consider exchange of massless particles.
$V_1^{(0)}$ has then a Coulombic expression (A4). The term $V_1^{(1)}$
can be calculated in perturbation theory from the two-particle exchange
diagrams [Eq. (2.10)]. One finds :
\begin{equation}
V_1^{(1)}\ =\ \bigl (V_1^{(0)}\bigr )^2\ ,
\end{equation}
with $V_1^{(0)}$ given in Eq. (A4). (The corresponding calculations will
be reported elsewhere.) As in the electromagnetic case, it is possible to
modify the wave function representation by means of the inverse of the
transformation (5.7), with $\widetilde U^{(0)}$ now replaced by $V_1^{(0)}/2\
$,
and to reach a Coulomb-gauge like representation. It can be checked that
in the resulting Hamiltonian all the quadratic terms in the potential
disappear. Also, the momentum dependent part of the effective potential
then coincides with that obtained by Olsson and Miller \cite{om}, who, with a
different method took into account the retardation effects. (For a
discussion see also Ref. \onlinecite{bgxz}.)\par
The foregoing results lead us to the general conclusion that, when
covariant propagators are used, it is necessary to evaluate the leading
contribution coming from the two-particle exchange diagrams, in order to
obtain the correct $O(c^{-2})$ effects. It is only in the Coulomb (or
Landau) gauge for the electromagnetic interaction or in a particular
representation of the wave function for the scalar interaction that the
above evaluation may be circumvented.\par
As far as the spin dependent terms are concerned, Hamiltonian (5.3) agrees
with the general expressions obtained previously in the literature
\onlinecite{hkmsbgmcb}.\par
Scalar potentials play also an important role in the description of
interquark confining interactions. Confining potentials, however, do not
correspond to simple diagrams of perturbation theory and therefore they
may display complicated or nontrivial energy dependences, other than
those of Eq. (5.1), implied by the underlying dynamics. These energy
dependences do not affect, to order $1/c^2\ $, the spin dependent part
of the Hamiltonian, but modify its momentum dependent part.\par
To analyze the structure of the Hamiltonian corresponding to a confining
scalar potential, we shall adopt for the latter a more general energy
dependence than that given in Eq. (5.1) by perturbation theory :
\begin{eqnarray}
&\ & V_1\ =\ \frac {1}{2M}\ \widetilde V_1^{(0)}\bigg \vert _{P_L = M}\ +\
\frac {(P_L - M)}{2M}\ \frac {\partial \widetilde V_1^{(0)}}{\partial P_L}
\bigg \vert _{P_L = M}\ +\ \frac {V_1^{(1)}}{2M^2}\ +\ O(c^{-6})\ ,\nonumber \\
&\ & \widetilde V_1^{(0)}\bigg \vert _{P_L = M}\ =\ V_1^{(0)}\ .
\end{eqnarray}
Here, $V_1^{(0)}$ is the nonrelativistic confining potential,
while the second term in the right-hand side
of the first equation takes into account the
leading energy dependence of the relativistic potential $V_1\ $. As in the
perturbative case, $V_1^{(1)}$ is assumed to be a quadratic function of
$V_1^{(0)}$ and presumably arises from the expansion of the relativistic
potential $V_1$ in terms of $V_1^{(0)}\ $.\par
The expression of the Hamiltonian is dependent on the wave function
representation, which is modified by transformations of the type (5.7).
The latter are equivalent to the introduction of pure gauge vector
potentials. These are described by the potentials $U_4$ and $T_4\ $, the
leading terms of which are parametrized as :
\begin{equation}
U_4\ =\ \frac {1}{2M}\ U^{(0)}\ ,\ \ \ \ T_4\ =\ \frac {1}{2M}\ T^{(0)}\ ;
\end{equation}
they satisfy the second of Eqs. (5.5) :
\begin{equation}
T^{(0)}\ =\ 2{\bf x^2}\ U^{(0)}{}'\ .
\end{equation}
Notice, however, that $U^{(0)}$ and $T^{(0)}$ may still depend on ratios of
masses, not present in perturbation theory.\par
As shown in subsection B, pure gauge vector potentials do not contribute
to the spin dependent terms (to order $1/c^2$).\par
We shall ignore in our analysis the short distance vector potential,
which has the structure of the electromagnetic potential, discussed in
subsection B, and does not affect the leading large distance properties
of the confining potential.\par
The resulting Hamiltonian takes the form (in the c.m. frame) :
\widetext
\begin{eqnarray}
H\ =&\ & \frac {{\bf p^2}}{2\mu}\ +\ V_1^{(0)}\ -\ \frac {1}{8}\
\bigl (\ \frac {1}{m_1^3} + \frac {1}{m_2^3}\ \bigr )\ ({\bf p^2})^2
\nonumber \\
&-& \frac {1}{4}\ \bigl (\ \frac {1}{m_1^2} + \frac {1}{m_2^2} -
\frac {1}{m_1 m_2}\ \bigl )\ \bigl [\ {\bf p^2}\ ,\ V_1^{(0)}\ \bigr ]_+\
+\ \frac {M}{4 m_1 m_2}\ \bigl [\ {\bf p^2}\ ,\
\frac {\partial \widetilde V_1^{(0)}}{\partial P_L}\ \bigr ]_+\nonumber \\
&-& \frac {1}{2m_1 m_2}\ \bigl [\ {\bf p^2}\ ,\ U^{(0)} + T^{(0)}\
\bigr ]_+\ +\ \frac {1}{m_1 m_2}\ \frac {T^{(0)}}{{\bf x^2}}\ {\bf L^2}
\nonumber \\
&-& \frac {\hbar ^2}{8}\ \bigl (\ \frac {1}{m_1^2} + \frac {1}{m_2^2}\
\bigr )\ \Delta V_1^{(0)}\ +\ \frac {\hbar ^2}{4 m_1 m_2}\ \Delta
(\ U^{(0)} - T^{(0)}\ )\nonumber \\
&+& \frac {1}{M}\ V_1^{(1)}\ +\ \frac {1}{2M}\ \bigl (V_1^{(0}\bigr )^2\ +\
V_1^{(0)}\ \frac {\partial \widetilde V_1^{(0)}}{\partial P_L}\ -\
\bigl (\ \frac {{\bf L.s_1}}{m_1^2} + \frac {{\bf L.s_2}}{m_2^2}\
\bigr )\ V_1^{(0)}{}'\ .\nonumber \\
\
\end{eqnarray}
\narrowtext
(The prime represents derivation with respect to ${\bf x^2}$ and $\Delta $
is the Laplacian; $[\ ,\ ]_+$ is the anticommutator.)\par
A possible comparison of this Hamiltonian with other expressions obtained
from quantum field theoretic calculations would permit the determination of its
various pieces  and, with reasonable assumptions, might lead to a
reconstitution of the covariant expression of the confining potential.\par
In this respect, we mention the work of Barchielli, Montaldi and Prosperi
\cite{bmpbpbbp}, who, continuing the Wilson loop approach to the confining
potential \cite{wadmbw}, as was developed earlier by Eichten and Feinberg
\cite{ef} and by Gromes \cite{dg}, obtained from QCD an expression for the
Hamiltonian to order $1/c^2$, including its momentum dependent part.\par
The complete reconstitution of the confining potential from the
comparison of the Hamiltonian (5.16) with that of Ref. \onlinecite{bmpbpbbp}
goes beyond the scope of the present paper and is left for future work. We
simply sketch here the qualitative features obtained from such a
comparison. As was already pointed out by Olsson and collaborators
\cite{lcooowow}, the Hamiltonian of Ref. \onlinecite{bmpbpbbp} favors
the interpretation of the underlying dynamics with the semiclassical flux
tube picture. Actually, this result is not very surprising, since one of
the initial hypotheses used in Ref. \onlinecite{bmpbpbbp}, namely the
dominance of the longitudinal color electric field, is itself motivated,
at least partly, from that picture.\par
The expression of the relativistic confining potential $V_1$ suggested by
the above comparison, turns out to be momentum dependent, typical of flux
tube models. This means that in $x$-space $V_1$ should be represented by
an integral operator (in the three-dimensional variable $x^T$). However,
in the classical case, the momentum dependent part is a smooth function,
and therefore, as a first approximation, one can use for it a mean value
approximation, thus representing $V_1$ by a local function in $x^T\ $.\par
We also notice that the momentum dependent part of the Hamiltonian of
Ref. \onlinecite{bmpbpbbp} does not vanish in the limit
$m_2 \rightarrow \infty \ $. Therefore, in this limit, particle 1 will
satisfy a Dirac equation with still a momentum dependent potential.\par
\pagebreak

\section{Summary and concluding remarks}

We showed that the two-fermion relativistic wave equations of Constraint
Theory can be reduced, by expressing the components of the $4\times 4$
matrix wave function in terms of one of the $2\times 2$ components, to
a single equation of the Pauli-Schr\"odinger type, valid for all sectors
of quantum numbers. This equation, which is relativistic invariant,
can be analyzed and solved with the usual tools of nonrelativistic
quantum mechanics - a feature which considerably simplifies the
treatment of two-fermion relativistic problems.\par
The interaction potentials that are present in this equation belong to
the general classes of scalar, pseudoscalar and vector interactions
and are calculable in perturbation theory from Feynman diagrams.
They are of the quasipotential type, in the sense that even in local
approximations, which we adopted throughout this work, they may still
depend on the c.m. energy of the system. This property allows one to
take into account leading contributions of nonlocal effects.\par
In the limit when one of the masses becomes infinite, the equation reduces
to the two-component form of the one-particle Dirac equation
with external static potentials.\par
The Hamiltonian of the system to order $1/c^2$ reproduces most the known
theoretical results obtained by other methods. Furthermore, in the
electromagnetic case, gauge invariance of the wave equation is checked
to this order, by considering the photon propagator in arbitrary
covariant gauges.\par
A last application was devoted to the analysis of the structure of the
Hamiltonian, to order $1/c^2\ $, in the case of confining interactions.
We emphasized here the role of the c.m. energy dependence of the
relativistic potential. We displayed, for confining scalar
potentials, the general expression of
the corresponding Hamiltonian, in the presence of pure gauge vector
potentials. This expression can be used for comparison with other
theoretical evaluations and reconstitution of the relativistic
confining potential.\par
The relativistic invariance of the reduced Pauli-Schr\"odinger type
equation also allows one to consider ultra-relativistic limits, which
play a crucial role in particle physics. \par
\pagebreak

\appendix

\section{The potentials from one-particle exchange diagrams}

In this appendix we display the relationships of the various potentials,
introduced in Sec. II and used throughout this paper, with the
propagators of the exchanged particles in perturbation theory.
According to formula (2.10), the three-dimensional reduction of a
propagator with mass $\mu $ results, in momentum space, from the operation :
\begin{equation}
\widetilde D(q^T,\mu )\ =\ D(q_L=0,q^T,\mu )\ ,
\end{equation}
or in $x$-space :
\begin{equation}
\widetilde D(x^T,\mu )\ =\ \int dx_L\ D(x,\mu )\ ,
\end{equation}
$D$ representing here the usual four-dimensional propagator.
For scalar particles and in lowest order of perturbation theory
the propagator is
$D(q,\mu )\ =\ i/(q^2 - \mu ^2 + i\epsilon )\ $.
\par
For scalar interactions one has :
\begin{equation}
V_1(x^T)\ =\ (-i)^2\ \frac {i}{2P_L}\ g_1 g_2\ \widetilde D(x^T,\mu)\ ,
\end{equation}
$g_1$ and $g_2$ being the coupling constants of the external particles
to the exchanged particle with mass $\mu $. In lowest order :
\begin{equation}
V_1(x^T)\ =\ -\frac {1}{2P_L}\ \frac {g_1 g_2}{4\pi }\
\frac {e^{\displaystyle -\mu r}}{r}\ ,\ \ \ \ r\ =\ \sqrt {-x^{T2}}\ .
\end{equation}
\par
For pseudoscalar interactions, the potential $V_3$ has the same type of
definitions as $V_1\ $.\par
For QED one has :
\begin{equation}
D_{\mu \nu }(k)\ =\ -\bigl (\ g_{\mu \nu } D(k,0)
- k_{\mu } k_{\nu } F(k)\ \bigr )\ ,
\end{equation}
where the gauge function $F$ is taken as a covariant function of $k\ $. Then :
\begin{eqnarray}
\widetilde D_{\mu \nu }(x^T)\ &=& -\bigl (\ g_{\mu \nu} \widetilde D(x^T,0) +
\partial_{\mu }^T \partial _{\nu }^T \widetilde F(x^T)\ \bigr )\nonumber \\
&=& -\bigl (\ g_{\mu \nu } \widetilde D +
2 g_{\mu \nu }^{TT} \dot {\widetilde F}
+ 4x_{\mu }^T x_{\nu }^T \ddot {\widetilde F}\ \bigr )\ ,
\end{eqnarray}
the dot operation being defined in Eq. (2.24).\par
By identification with Eq. (2.20) we find :
\begin{eqnarray}
U_4 &=& V_2\ +\ \widetilde U\ ,\nonumber \\
T_4 &=& 2x^{T2}\ \dot {\widetilde U}\ .
\end{eqnarray}
\par
For particles with opposite charges we have :
\begin{equation}
V_2\ =\ -(-i)^2\ \frac {i}{2P_L}\ e^2\ \widetilde D(x^T,0)\ .
\end{equation}
In lowest order, $V_2$ becomes :
\begin{equation}
V_2\ =\ \frac {1}{2P_L}\ \frac {\alpha }{r}\ ,\ \ \
r\ =\ \sqrt {-x^{T2}}\ ,\ \ \ \alpha \ =\ \frac {e^2}{4\pi }\ .
\end{equation}
(Notice that the sign of the charge of the antiparticle is taken
into account by the matrices $\gamma _2^{\nu }$ of Eq. (2.20) which
act on the wave function from the right. In the nonrelativistic
limit, the matrix $\gamma _{2L}$ has eigenvalue $-1$ .)\par
For covariant gauges described by a parameter $\xi $, we have in
lowest order :
\begin{equation}
\widetilde D_{\mu \nu }(x^T)\ =\ -\bigl (\ g_{\mu \nu } - g_{\mu \nu }^{TT}\
\frac {\xi }{2} +  \frac {x_{\mu }^T x_{\nu }^T} {x^{T2}}\
\frac {\xi }{2}\ \bigr )\ \widetilde D(x^T,0)\ .
\end{equation}
The Feynman gauge is obtained with $\xi =0$ and the Landau gauge
with $\xi =1\ $.\par
For vector interactions resulting from the exchange of a vector particle
with mass $\mu \ $, the function $\widetilde F$ in Eq. (A6) should be replaced
by $\widetilde D/\mu ^2\ $.\par
As a final remark, we emphasize that relationships like (A3) and (A8)
between the potentials and the exchanged particle propagators are {\it not}
equivalent to the instantaneous approximation. The perturbation theory
expansion of Eq. (2.10) always reproduces for the potential corresponding
to a one-particle exchanged diagram (including vacuum polarization
effects) a local expression in $x^T$, together with the energy factor
$1/(2P_L)\ $. In the present formalism, the instantaneous approximation
produces a nonlocal potential due to the presence of the momentum
operator $p^{T2}$ in it \cite{s2}. On the other hand, the instantaneous
approximation does not provide the correct infinite mass limit, neither
the correct $1/c^2$ expansion.\par
\pagebreak

\section{Useful formulas for the $\gamma $-matrices and the spin and
orbital angular momentum operators}

In this appendix we present formulas for the $\gamma $-matrices and
the spin and orbital angular momentum operators that are used during the
reduction operations from Eqs. (2.21) to Eq. (3.20). We also exhibit the
relationships between the components $\psi _1\ ,\ \psi _2 $ and $\psi _4 $
with $\psi _3 $ [Eq. (3.1)].\par
The transition from Eqs. (2.21) to Eqs. (2.23) is obtained with the use of
the following properties of the $\gamma $-matrices :
\widetext
\begin{eqnarray}
e^{\displaystyle -\lambda \overline V}\ (\gamma _{1\alpha }^T \pm
\gamma _{2\alpha }^T)\ e^{\displaystyle \lambda V}\ =\
&e^{\displaystyle \pm 2\lambda U_4}&\ \big [\ (\gamma _{1\alpha }^T
\pm \gamma _{2\alpha }^T)\nonumber \\
&+& (\gamma _1^T.x^T \pm \gamma _2^T.x^T)\ \frac {1}{x^{T2}}\
(e^{\displaystyle \pm 2\lambda T_4} - 1)\ x_{\alpha }^T\ \big ]\ ,
\end{eqnarray}
\narrowtext
where $V$ is defined in Eq. (2.20) and $\overline V$ is defined as :
\begin{equation}
\overline V\ =\ 2V_1 - V\ .
\end{equation}
Equations (B1) are easily established by differentiating with respect to
the parameter $\lambda \ $.\par
We notice that all the bilinear forms of the $\gamma $-matrices present in
$V$ [Eq. (2.20] commute with each other.\par
Similarly one establishes the following relation giving the transformation
law of $p^T$ under the action of the exponential of $V$ :
\widetext
\begin{eqnarray}
e^{\displaystyle -\lambda V}\ p_{\alpha }^T\ e^{\displaystyle \lambda V}\
=\ p_{\alpha }^T\ &+&\
2i\hbar \lambda \ \big [\ \dot V_1\ +\ \gamma _{15} \gamma _{25} \dot V_3
\nonumber \\
&+&\ \gamma _1^{\mu } \gamma _2^{\nu }\ \big (\ g_{\mu \nu }^{LL}\dot V_2\
+\ g_{\mu \nu }^{TT}\dot U_4\ +\ \frac {x_{\mu }^T x_{\nu }^T}{x^{T2}}
\dot T_4\ \big )\ \big ]\ x_{\alpha }^T\nonumber \\
&+&\ \frac {i\hbar }{2x^{T2}}\ \big [\ (\sinh 2\lambda T_4)\
(\gamma _{1\alpha }^T \gamma _2^T.x^T + \gamma _{2\alpha }^T \gamma _1.x^T)
\nonumber \\
&\ &\ \ \ \ \ \ +\ i(\cosh 2\lambda T_4 - 1)\ (\sigma _{1\alpha \beta }^{TT}
+ \sigma _{2\alpha \beta }^{TT})\ x^{T\beta }\nonumber \\
&\ &\ \ \ \ \ \ -\ 2(\sinh 2\lambda T_4)\
\frac {1}{x^{T2}}\ \gamma _1.x^T \gamma _2.x^T x_{\alpha }^T\ \big ]\ ,
\end{eqnarray}
\narrowtext
where the dot operation is defined in Eq. (2.24).\par
The transition from Eq. (3.17) to Eq. (3.20) uses the following properties
of the spin operators, defined in Eqs. (3.3)-(3.5) :
\widetext
\begin{eqnarray}
e^{\displaystyle -A+wB}\ (W_{1S} \pm W_{2S}).p\ e^{\displaystyle A+wB}\ &=&\
e^{\displaystyle \mp 2B}\ \big [\ (W_{1S} \pm W_{2S}).p\nonumber \\
&\ &+\ 2i\hbar\ (\dot A \mp 2\dot B - w\dot B)\ (W_{1S} \pm W_{2S}).x^T\
\big ]\ ,\nonumber \\
e^{\displaystyle -A+wB}\ (W_{1S} \pm W_{2S})_{\alpha }\
e^{\displaystyle A+wB}\ &=&\ e^{\displaystyle \mp 2B}\ (W_{1S} \pm W_{2S})
_{\alpha }\ ,\nonumber \\
e^{\displaystyle w_{12} C}\ (W_{1S} \pm W_{2S})_{\alpha }\
e^{\displaystyle w_{12} C}\ &=&\ ( W_{1S} \pm W_{2S} )_{\alpha }\nonumber \\
&\ &-\ \frac {1}{x^{T2}}\ (1-e^{\displaystyle \mp 2C})\
(W_{1S} \pm W_{2S}).x\ x_{\alpha }^T\ ,\nonumber
\end{eqnarray}
\begin{eqnarray}
e^{\displaystyle -w_{12} C}\ p_{\alpha }^T\ e^{\displaystyle w_{12} C}\ =\
p_{\alpha }^T\
+\ \frac {i\hbar }{4 x^{T2}}\ \big (\frac {2}{\hbar P_L}\big )^2\
\bigg \{\ (1-e^{\displaystyle -2C})\ W_{S \alpha }\ W_S.x^T \nonumber \\
+\ (1-e^{\displaystyle 2C})\ (W_{1S} - W_{2S})_{\alpha }\
(W_{1S} - W_{2S}).x^T\
+\ \hbar ^2  P^2\ (1-\cosh 2C)\ x_{\alpha }^T
\nonumber \\
-\ 4(\sinh 2C)\ \frac {W_{1S}.x^T W_{2S}.x^T}{x^{T2}}\ x_{\alpha }^T\
+\ 8 \dot C\ W_{1S}.x^T W_{2S}.x^T\ x_{\alpha }^T\ \bigg \}\ .\nonumber \\
\end{eqnarray}
\narrowtext
Here $A,\ B,\ C$ are functions of $x^{T2}$ (and eventually of $P^2$).\par
The operator $w$ can be decomposed along projectors,
$w_0$ and $w_1$, on spin 0 and spin 1 states, respectively,
which also satisfy with $w_{12}$ the following relations :
\begin{eqnarray}
w\ &=&\ 3w_0 - w_1\ ,\ \ \ w_0\ =\ \frac {1}{4} (1+w)\ ,\ \ \ w_1\ =\
\frac {1}{4} (3-w)\ ,\nonumber \\
w(1-w_{12})\ &=&\ -(1-w_{12})\ ,\ \ \ w(1+w_{12})\ =\ 2w + 1 - w_{12}\ ,
\nonumber \\
w_0\ \frac {1}{2} (1-w_{12})\ &=&\ 0\ ,\ \ \ w_0\ \frac {1}{2}(1+w_{12})\ =\
w_0
\ , \nonumber \\
w_1\ \frac {1}{2} (1-w_{12})\ &=&\ \frac {1}{2} (1-w_{12})\ ,\ \ \
w_1\ \frac {1}{2} (1+w_{12})\ =\ \frac {1}{2} (1+w_{12}) - w_0\ ,\nonumber \\
e^{\displaystyle -w_{12} C}\ &=&\ \frac {1}{2} (1-w_{12})\
e^{\displaystyle C}\ +\ \frac {1}{2} (1+w_{12})\ e^{\displaystyle -C}\ ,
\nonumber \\
e^{\displaystyle A + wB}\ &=&\ w_0\ e^{\displaystyle A + 3B}\ +\
w_1\ e^{\displaystyle A - B}\ .\nonumber \\
\end{eqnarray}
\par
We also list here several useful formulas (in the c.m. frame) involving the
momentum, the spin and the orbital angular momentum operators
[Eqs. (3.18)] :
\widetext
\begin{mathletters}
\begin{eqnarray}
\big [\ {\bf p^2}\ ,\ f\ \big ]\ &=&\ -\ \big (\ 4i\hbar f'\ {\bf x.p}\ +\
6\hbar ^2 f'\ +\ 4\hbar ^2 {\bf x^2} f''\ \big )\ ,\\
\big [\ {\bf p^2}\ ,\ S_{12}\ g\ \big ]\ &=&\ -\ S_{12}\ \big (\ 4i\hbar g'\
{\bf x.p}\ +\ 6\hbar ^2 g'\ +\ 4\hbar ^2 {\bf x^2} g''\ \big )\nonumber \\
&\ &\ +\ \frac {1}{{\bf x^2}}
\big (\ 4S_{12}\ {\bf L.S}\ -\ 8{\bf s_1.s_2}\ +\ 6\hbar ^2 S_{12}
\ \big )\ g\
\end{eqnarray}
\end{mathletters}
\narrowtext
($f$ and $g$ are functions of ${\bf x^2}$ and eventually of $P^2$
and the prime operations are defined in Eqs. (3.21)),\par
\widetext
\begin{eqnarray}
&\ &{\bf (s_1\pm s_2).x\ (s_1\pm s_2).p}\ =\ \frac {1}{2}(1\pm S_{12})\
\big (\ {\hbar ^2} {\bf x.p} + {ih} {\bf L.S}\ \big )\ ,
\nonumber \\
&\ &{\bf (s_1\mp s_2).x\ (s_1\pm s_2).p}\ =\ \frac {i\hbar}{2}
(1\mp S_{12})\ {\bf L.(s_1-s_2)}\ ,\nonumber \\
&\ &\big [\ S_{12}\ ,\ {\bf L.(s_1-s_2)}\ \big ]_+\ =\ 0\ ,\ \ \ \ \
\big [\ S_{12}\ ,\ \big [\ S_{12}\ ,\ {\bf L.(s_1-s_2)}\ \big ]\ \big ]_+\
=\ 0\ ,\nonumber \\
&\ &w\ (1+S_{12})\ {\bf L.(s_1-s_2)}\ =\ -(1+S_{12})\ {\bf L.(s_1-s_2)}\ ,
\nonumber \\
&\ &w\ (1-S_{12})\ {\bf L.(s_1-s_2)}\ =\ 3(1-S_{12})\ {\bf L.(s_1-s_2)}\ ,
\nonumber \\
&\ &(1\pm S_{12})\ {\bf L.(s_1-s_2)}\ (1\pm S_{12})\ =\ 0\ ,\nonumber \\
&\ &(1\pm S_{12})\ {\bf L.(s_1-s_2)}\ (1\mp S_{12})\ =\ 2{\bf L.(s_1-s_2)}\
\pm \ \big [\ S_{12}\ ,\ {\bf L.(s_1-s_2)}\ \big ]\ ,\nonumber \\
&\ &\big [\ S_{12}\ ,\ {\bf L.S}\ \big ]_+\ =\ 2{\bf S^2} -3\hbar ^2 -
3\hbar ^2 S_{12}\ ,\nonumber \\
&\ &(1+S_{12})\ {\bf L.S}\ (1+S_{12})\ =\ -2\hbar ^2 (1+S_{12})\ ,\nonumber \\
&\ &(1-S_{12})\ {\bf L.S}\ (1-S_{12})\ =\ 4\hbar ^2 (1+S_{12})\ -\
4{\bf S^2}\ ,\nonumber \\
&\ &(1\pm S_{12})\ {\bf L.S}\ (1\mp S_{12})\ =\ 2\ \big (\ {\bf L.S +S^2}
- \frac {{\bf (S.x)^2}}{\hbar ^2 {\bf x^2}}\ \big )\ \pm\
\big [\ S_{12}\ ,\ {\bf L.S}\ \big ]\ ,\nonumber \\
&\ &\big [\ S_{12}\ ,\ \big [\ S_{12}\ ,\ {\bf L.S}\ \big ]\ \big ]_+\ =\ 0\ ,
\ \ \ \ \ S_{12}\ \big [\ S_{12}\ ,\ {\bf L.S}\ \big ]\ =\
2\ \big (\ {\bf L.S + S^2 - \frac {(S.x)^2}{x^2}}\ \big )\ \nonumber \\
\
\end{eqnarray}
\narrowtext
($[\ ,\ ]_+$ is the anticommutator),\par
\widetext
\begin{eqnarray}
4\ {\bf s_1.p\ s_2.p}\ =\ \hbar ^2\ S_{12}\ {\bf p^2}\ +\
\frac {i\hbar }{{\bf x^2}}\ \big [\ S_{12}\ ,\ {\bf L.S}\ \big ]\ {\bf x.p}
\ -\ \frac {1}{{\bf x^2}}\ \big [\ S_{12}\ ,\ {\bf L.S}\ \big ]\ {\bf L.S}\ .
\nonumber \\
\
\end{eqnarray}
\narrowtext
\par
\pagebreak
The relation expressing $\psi _4$ in terms of $\psi _3$ is :
\widetext
\begin{eqnarray}
\frac {1}{4} MP_L\ \psi _4\ =&\ & \bigg [\ \frac {1}{2} (1+S_{12})\
\frac {P^2}{4}\ e^{\displaystyle 2(V_2-V_1)}\ +\ \frac {1}{2}(1-S_{12})\
\frac {M^2}{4}\ e^{\displaystyle 2(V_1-V_2)}\ \bigg ]\ \psi _3\nonumber \\
&-&\ e^{\displaystyle -[4U_4 + 2(V_1+V_2) +2h]}\
\bigg \{\ \frac {1}{2\hbar ^2 {\bf x^2}}\ \big [\ S_{12}\ ,\ {\bf L.S}\ \big ]\
(\ -i\hbar \ {\bf x.p +\ L.S}\ )\nonumber \\
&+&\ \frac {i\hbar }{2\hbar ^2 {\bf x^2}}\ (1-e^{\displaystyle -2T_4})\
\big [\ S_{12}\ ,\ {\bf L.S\ \big ]\ x.p}\nonumber \\
&-&\ \big [\ 2( V_1' - V_2' + V_3' + U_4' )\ e^{\displaystyle -2T_4}\
-\ \frac {1}{2 {\bf x^2}} (1-e^{\displaystyle -2T_4})\ \big ]\
\big ( {\bf L.S + S^2 - \frac {(S.x)^2}{x^2}}\big )\nonumber \\
&+&\ \big [\ T_4'\ e^{\displaystyle -2T_4}\ +\ \frac {3}{4{\bf x^2}}\
(1-e^{\displaystyle -2T_4})\ \big ]\ \big [\ S_{12}\ ,\ {\bf L.S}\ \big ]\
\bigg \}\ \psi _3\ .\nonumber \\
\
\end{eqnarray}
\narrowtext
\par
Notice that the terms in the curly brackets disappear in the sectors with
quantum numbers $j=\ell \ $.\par
The relation expressing $\psi _3$ in terms of $\psi _4$ can be deduced
from Eq. (B9) with the following substitutions : $\psi _3
\leftrightarrow \psi _4\ ,\ M^2 \leftrightarrow P^2\ ,\ V_1
\leftrightarrow V_2$ and a global change of sign in front of the curly
brackets.\par
Equations (3.14), which express $\psi _1$ and $\psi _2$ in terms of
$\psi _3 \ $, can be rewritten in more explicit forms when potential $V$
is given by Eq. (2.20) :
\widetext
\begin{mathletters}
\begin{eqnarray}
\psi _1\ =\ \frac {2}{\hbar P_L}&\times & e^{\displaystyle -2(U_4 + V_2 + h)}\
\bigg \{\ {\bf (s_1-s_2).p}\ -\ \frac {1}{{\bf x^2}}\
(1-e^{\displaystyle -2T_4})\ {\bf (s_1-s_2).x\ x.p}\nonumber \\
&+& \frac {i\hbar }{2{\bf x^2}}\ (1-e^{\displaystyle -2T_4})\
(4-\frac {2{\bf S^2}}{\hbar ^2})\ {\bf (s_1-s_2).x}\nonumber \\
&-& 2i\hbar \ \big (\ V_1' - V_2' + V_3' + U_4' - T_4' -
\frac {2{\bf S^2}}{\hbar ^2}\ U_4'\ \big )\ e^{\displaystyle -2T_4}\
{\bf (s_1-s_2).x}\nonumber \\
&-& \frac {(m_1^2 - m_2^2)}{M^2}\ e^{\displaystyle -4V_1}\
\bigg [\ {\bf S.p}\ -\ \frac {1}{{\bf x^2}} (1-e^{\displaystyle -2T_4})\
{\bf S.x\ x.p}\nonumber \\
&+& \frac {i\hbar }{{\bf x^2}}\ (1-e^{\displaystyle -2T_4})\ {\bf S.x}
\nonumber \\
&+& 2i\hbar \ \big (\ V_1' - V_2' + V_3' + U_4' + T_4'\ \big )\
e^{\displaystyle -2T_4}\ {\bf S.x}\ \bigg ]\ \bigg \}\ \psi _3\ ,\nonumber \\
\
\end{eqnarray}
\pagebreak
\begin{eqnarray}
\psi _2\ =\ \frac {2}{\hbar M}&\times & e^{\displaystyle -2(U_4 + V_1 +h)}\
\bigg \{\ -\ {\bf S.p} + \frac {1}{{\bf x^2}}\ (1-e^{\displaystyle -2T_4})\
{\bf S.x\ x.p}\nonumber \\
&-& \frac {i\hbar }{{\bf x^2}}\ (1-e^{\displaystyle -2T_4})\ {\bf S.x}
\nonumber \\
&-& 2i\hbar \ \big (\ V_1' - V_2' + V_3' + U_4' + T_4'\ \big )\
e^{\displaystyle -2T_4}\ {\bf S.x}\nonumber \\
&+& \frac {(m_1^2 - m_2^2)}{P^2}\ e^{\displaystyle -4V_2}\
\bigg [\ {\bf (s_1-s_2).p}\ -\ \frac {1}{{\bf x^2}} (1-e^{\displaystyle -2T_4})
\ {\bf (s_1-s_2).x\ x.p}\nonumber \\
&+& \frac {i\hbar }{2{\bf x^2}}\ (1-e^{\displaystyle -2T_4})\
(4-\frac {2{\bf S^2}}{\hbar ^2})\ {\bf (s_1-s_2).x}\nonumber \\
&-& 2i\hbar \ \big (\ V_1' - V_2' + V_3' + U_4' - T_4' -
\frac {2{\bf S^2}}{\hbar ^2}\ U_4'\ \big )\ e^{\displaystyle -2T_4}\
{\bf (s_1-s_2).x}\ \bigg ]\ \bigg \}\ \psi _3\ .\nonumber \\
\
\end{eqnarray}
\end{mathletters}
\narrowtext
\par
\pagebreak

\section{Cancellation of high order ladder diagrams in the infinite mass
limit}

We show in this appendix that in the infinite mass limit
($m_2 \rightarrow \infty $) the high order ladder and crossed ladder
diagrams are cancelled by the ``constraint diagrams'', so that the
dominant contribution comes from the one-particle exchange diagrams only.
\par
We shall ignore the contributions of vertex corrections, which yield
nonlocal terms in the potential, even in the one-particle problem.
For simplicity, internal fermion propagators will be considered as free,
but this assumption does not restrict the generality of the result.
Vacuum polarization may be included in the exchanged particle
propagator. For definiteness, we consider the case of QED (with coupling
constants $e_1$ and $e_2$ for fermions 1 and 2, respectively), but the
derivation can be repeated with other types of interaction as well.
Mass of the exchanged particle is irrelevant.\par
In the approximation of ladder and crossed ladder diagrams, the
off-mass shell (amputated) scattering amplitude can be decomposed as :
\begin{equation}
T\ =\ \sum _{n=1}^{\infty } T_n\ ,
\end{equation}
where $T_n$ is the partial amplitude corresponding to $n$ exchanged
photons. A similar decomposition also holds for the constrained
amplitude $\widetilde T$ (2.10).\par
The iteration of Eq. (2.10) yields for the potential the expression :
\begin{equation}
\widetilde V\ =\ \widetilde T\ \sum _{p=0}^{\infty }
\bigl ( G_0 \widetilde T \bigr )^p \ \equiv \
\sum _{n=0}^{\infty } \widetilde V_n\ ,
\end{equation}
where $\widetilde V_n$ is that part of the potential which comes from
the contributions of $n$ exchanged photons. The first two terms of
$\widetilde V$ are :
\begin{eqnarray}
\widetilde V_1 &=& \widetilde T_1\ ,\\
\widetilde V_2 &=& \widetilde T_2\ +\ \widetilde T_1 G_0 \widetilde T_1\ .
\end{eqnarray}
The corresponding Feynman diagrams are represented in Fig. 1. While
$\widetilde T_2$ contains the two usual Feynman diagrams of two
exchanged photons, the last term in Eq. (C4) yields the ``constraint
diagram'' (third diagram in Fig. 1b).\par
The expression of $\widetilde V_1$ was given in Appendix A (notice that
to this order the potentials $\widetilde V$ and $V$ [Eq. (2.17)]
coincide); in the limit $m_2 \rightarrow \infty \ $, $\widetilde V_1$
behaves as $O(1/m_2)$. We now concentrate on the behavior of
$\widetilde V_2$.\par
In $\widetilde T$, the external momenta are submitted to the constraint
(2.11) and, therefore, satisfy the relations :
\begin{eqnarray}
p_1 + p_2 &=& p'_1 + p'_2\ =\ P\ ,\nonumber \\
p_1^2 - p_2^2 &=& {p'}_1^2 - {p'}_2^2\ =\ m_1^2 - m_2^2\ ,\ \ \
p'_{1L}=p_{1L}\ , \ \ p'_{2L}=p_{2L}\ .
\end{eqnarray}
In the limit $m_2 \rightarrow \infty $, one has :
\begin{eqnarray}
p_{20} &=& m_2 + O\bigl (\frac {1}{m_2}\bigr )\ ,\ \ \
P_0\ =\ m_2 + p_{10} + O\bigl (\frac {1}{m_2}\bigr )\ ,\nonumber \\
p_{10} &-& p'_{10}\ =\ p'_{20} - p_{20}\ =\
O\bigl (\frac {\vert {\bf P}\vert }{m_2}\bigr )\ .
\end{eqnarray}
\par
The expression of $\widetilde T_2$ is :
\widetext
\begin{eqnarray}
\widetilde T_2 &=& (-i)^4 (e_1e_2)^2\ \bigl (\frac {i}{2P_L}\bigr ) \int
\frac {d^4k_1}{(2\pi )^4} \ D_{\mu _1 \nu _1} (k_1)\
D_{\mu _2 \nu _2} (p_1 - p'_1 - k_1)\nonumber \\
&\ & \ \ \ \ \times
\gamma _{1\mu _1} S_1 (p_1-k_1) \gamma _{1\mu _2}\
\bigl [\ \gamma _{2\nu _1} S_2 \bigl (-(p_2+k_1)\bigr ) \gamma _{2\nu _2} +
\gamma _{2\nu _2} S_2 \bigl (-(p'_2-k_1)\bigr ) \gamma _{2\nu _1}\ \bigr ]\ ,
\end{eqnarray}
\narrowtext
where the photon propagator is considered in an arbitrary covariant
gauge [Eq. (A5)].\par
In the limit $m_2 \rightarrow \infty$ one has \cite{b,ng} :
\widetext
\begin{eqnarray}
&\ &\bigl [\ \gamma _{2\nu _1} S_2 \bigl (-(p_2+k_1)\bigr )\gamma _{2\nu _2} +
\gamma _{2\nu _2}S_2 \bigl (-(p'_2-k_1)\bigr )\gamma _{2\nu _1}\ \bigr ]
\nonumber \\
&\ & \ \ \ \ \ \ \ \ \ = \
\bigl [\ \gamma _{2\nu _1}\frac {i}{-\gamma _2.(p_2+k_1) - m_2 +
i\epsilon } \gamma _{2\nu _2} + \gamma _{2\nu _2}\frac {i}
{-\gamma _2.(p'_2-k_1) - m_2 + i\epsilon } \gamma _{2\nu _1} \ \bigr ]
\nonumber \\
&\ & \ \ \ \ \ \ \ \ \ = \
(-1)^2 \delta _{\nu _1 0} \delta _{\nu _2 0}\ 2\pi \delta (k_{10})\
+\ O\bigl (\frac {1}{m_2}\bigr )\ ,
\end{eqnarray}
\narrowtext
where the eigenvalue $-1$ has been used for the matrix $\gamma _{20}$
(antifermion). (Because of the factor $\delta (k_{10})$ and Eqs. (C6), the
gauge part of the photon propagator disappears in Eq. (C7) at leading
order.)\par
On the other hand, in the iteration term $\widetilde T_1 G_0 \widetilde T_1$
[Eq. (C4)] the integration concerning $G_0$ is three-dimensional after the
use of the constraint (2.11) with the fermion momenta $(p_1 + k_1)$
and $(p_2 - k_1)$. This yields the condition $k_{1L}=0\ $. Hence, one
obtains :
\widetext
\begin{eqnarray}
\widetilde T_1 G_0 \widetilde T_1 &=& (-i)^4 (e_1e_2)^2\
{\bigl (\frac {i}{2P_L}\bigr )}^2 \int \frac {d^4k_1}{(2\pi )^4}\
2\pi \delta (k_{1L})\ D_{\mu _1 \nu _1} (k_1)\ D_{\mu _2 \nu _2}
(p_1 - p'_1 - k_1)\nonumber \\
&\ & \ \ \ \ \ \ \times
\gamma _{1\mu _1} S_1(p_1-k_1) \gamma _{1\mu _2}\ \gamma _{2\nu _1}
S_2 \bigl (-(p_2+k_1)\bigr ) \gamma _{2\nu _2}\ H_0(p_2+k_1)\ .
\end{eqnarray}
\narrowtext
Notice that because of the constraint (2.11) $H_0(p_1-k_1)=H_0(p_2+k_1)$
[Eq. (2.13)]. One also has :
\begin{equation}
S_2 \bigl (-(p_2+k_1)\bigr )\ H_0(p_2+k_1)\ =\
i\bigl (\ -\gamma _2.(p_2+k_1) + m_2\ \bigr )\ .
\end{equation}
\par
In the limit $m_2 \rightarrow \infty \ $, this term, multiplied by the
matrices $\gamma _{2\nu _1}$ and $\gamma _{2\nu _2}\ $, behaves as
$(-1)^2 \delta _{\nu _1 0} \delta _{\nu _2 0} (i2m_2)$ and the factor
$2m_2$ cancels one of the factors $2P_L$ of the denominator of the
right-hand side of Eq. (C9). Furthermore, in the same limit
$\delta (k_{1L})\rightarrow \delta (k_{10})$ and, because of the
additional factor $i^2$, the iteration term
$\widetilde T_1 G_0 \widetilde T_1$ cancels, at leading order, the
amplitude term $\widetilde T_2$ in Eq. (C4). Hence, $\widetilde V_2$
behaves, when $m_2 \rightarrow \infty \ $, as $O(1/m_2^2)\ $.\par
The above calculations can be repeated at higher orders of perurbation
theory. For the diagrams with $n$ exchanged photons, $\widetilde V_n$
is the sum of the partial amplitude $\widetilde T_n$ and the iteration
terms coming from $\widetilde V G_0 \widetilde T\ $; it can be written
in the form :
\begin{equation}
\widetilde V_n\ =\ \widetilde T_n\ +\ \sum _{p=1}^{n-1}\
\sum _{r_1 + \cdots + r_{p+1} =n} \widetilde T_{r_1} G_0
\widetilde T_{r_2} G_0 \cdots \widetilde T_{r_p} G_0 \widetilde T_{r_{p+1}}\ ,
\end{equation}
where, in the generic term of the sum, the constraint factor $G_0$
appears $p$ times. A typical diagram, where $p=2$, is shown in Fig. 2.\par
Putting aside all factors which do not play an essential role, an
amplitude $\widetilde T_r\ $, that contains all ladder and crossed ladder
type Feynman diagrams with $r$ exchanged photons, behaves in the limit
$m_2 \rightarrow \infty $ as \cite{b,ng} :
\begin{equation}
\widetilde T_r\ \sim \ (-i)^{2r} (-e_1e_2)^r\ \bigl (\frac {i}{2P_L}\bigr )
\ (2\pi )^{r-1}\ \delta (k_{10}) \cdots \delta (k_{r-10})\ ,
\end{equation}
while the occurrence of a constraint factor $G_0$ just after
$\widetilde T_r$ provides the contribution :
\begin{equation}
G_0\ \sim \ (i2m_2)\ 2\pi \delta (k_{r0})\ .
\end{equation}
\par
It can be checked , using these results, that a term of the iteration
sum in Eq. (C11), containing $p$ factors $G_0\ $, behaves as :
\begin{equation}
\widetilde T_{r_1} G_0 \widetilde T_{r_2} G_0 \cdots \widetilde T_{r_p}
G_0 \widetilde T_{r_{p+1}}\ \sim \ (-1)^p \widetilde T_n\ .
\end{equation}
\par
Furthermore, the $p$ factors $G_0$ may appear in such a term in
$\left (_{\ p}^{n-1}\right )$
different configurations. The behavior of $\widetilde V_n$ is then :
\begin{equation}
\lim _{m_2 \rightarrow \infty } m_2 \widetilde V_n\ =\
\lim _{m_2 \rightarrow \infty } m_2 \widetilde T_n\
\bigl [\ 1 + \sum _{p=1}^{n-1} (-1)^p \left (_{\ p}^{n-1}\right )\
\bigr ]\ =\ 0\ ,\ \ \ \ n\ge 2\ .
\end{equation}
\par
Therefore, $\widetilde V_n$ behaves as $O(1/m_2^2)$ and the dominant
contribution to $\widetilde V$ [Eq. (C2)] comes solely from the
one-particle exchange diagram, which behaves as $O(1/m_2)\ $.
A similar result is also obtained in other approaches, based on a
three-dimensional formulation of the two-body bound state problem
\cite{fh,rta}.
\par
\pagebreak

\begin{figure}
\caption{The diagrams contributing to $\widetilde V_1\ $, (a), and
$\widetilde V_2\ $, (b) [Eqs. (C2)-(C4)]. The third diagram in (b)
is the ``constraint diagram''.}
\end{figure}

\begin{figure}
\caption{A typical ``constraint diagram'' of high order [Eq. (C11)],
where the constraint factor $G_0$ [Eq. (2.12)] appears twice (boxes
with a cross). The boxes without a cross represent ordinary ladder
and crossed ladder type Feynman diagrams (contributing to the partial
amplitudes $\widetilde T_r$ in Eq. (C11)).}
\end{figure}

\end{document}